\documentclass[codesnippet]{jss}
\pdfoutput=1
\usepackage{amsmath,amsthm,amsfonts,amssymb,natbib,enumitem}
\newtheorem{alg}{Algorithm}
\author{Panagiotis Papastamoulis\\University of Piraeus}
\title{\pkg{label.switching}: An \proglang{R} Package for Dealing with the Label Switching Problem in MCMC Outputs}

\Plainauthor{Panagiotis Papastamoulis} 
\Plaintitle{\pkg{label.switching}: An R Package for dealing with the Label Switching problem in MCMC outputs} 
\Shorttitle{\pkg{label.switching}: Relabelling MCMC outputs} 

\Abstract{Label switching is a well-known and fundamental problem in Bayesian estimation of mixture or hidden Markov models. In case that the prior distribution of the model parameters is the same for all states, then both the likelihood and posterior distribution are invariant to permutations of the parameters. This property makes Markov chain Monte Carlo (MCMC) samples simulated from the posterior distribution non-identifiable. In this paper, the \pkg{label.switching} package is introduced. It contains one probabilistic and seven deterministic relabelling algorithms in order to post-process a given MCMC sample, provided by the user. Each method returns a set of permutations that can be used to reorder the MCMC output. Then, any parametric function of interest can be inferred using the reordered MCMC sample. A set of user-defined permutations is also accepted, allowing the researcher to benchmark new relabelling methods against the available ones.
}
\Plainkeywords{Label switching, mixture models, hidden Markov, MCMC, R} 

\Address{
  Panagiotis Papastamoulis\\
  Department of Statistics and Insurance Science\\
  University of Piraeus\\
  Greece\\
  E-mail: \email{papapast@yahoo.gr}
}

\begin{document}


\section[Introduction]{Introduction}

Mixture and hidden Markov models are a powerful tool for modelling a wide range of phenomena and they have been extremely useful in many fields \citep{McLachlan:00, Fruhwirth:06}. Such applications include the presence of unobserved heterogeneity in the studied population or the approximation of unknown distributions, after deciding the proper number of latent states (or components). The complex nature of such models can be simplified by decomposing the model into simpler structures using latent (unobserved) variables. Data augmentation \citep{tanner} is a standard technique exploited both by the EM algorithm \citep{Dempster:77} as well as the Gibbs sampler \citep{gelfand}. 

Under a Bayesian perspective, MCMC estimation of the posterior distribution of such models is quite straightforward. Gibbs sampling enables the simulation of a Markov chain from the joint posterior distribution of model parameters and latent variables. Nevertheless, the likelihood of such models is invariant to permutations of the components' labels and this property gives rise to the label switching phenomenon \citep{Redner, jasra}. It is well known that the presence of the label switching phenomenon in an MCMC sample serves as a necessary condition for the convergence of the chain to the target distribution. On the other hand, the presence of the phenomenon complicates the  posterior inference.

Early attempts to solve the label switching were focused on the use of suitable identifiability constraints \citep{diebolt, Richardson:97, Fruhwirth:01}. However, it is not always possible to find such constraints and in most cases a single identifiability constraint will not respect the geometry of the posterior distribution. For these reasons, a variety of different relabelling algorithms has been proposed.  

The purpose of this study is to introduce the  \pkg{label.switching} package, available from the Comprehensive \proglang{R} Archive Network at \url{http://CRAN.R-project.org/package=label.switching}, which can be used in order to deal with the label switching problem using various algorithms that have been proposed to the related literature. More specifically, the \pkg{label.switching} package consists of eight relabelling methods: ordering constraints, the Kullback-Leibler
based algorithm \citep{stephens}, the pivotal reordering algorithm \citep{Marin:05, Marin:07}, the default and iterative versions of ECR algorithm \citep{Papastamoulis:10,Rodriguez,Papastamoulis:13,Papastamoulis:14},  the probabilistic relabelling algorithm  \citep{sperrin} and the data-based algorithm \citep{Rodriguez}. 

In many instances, it is required to draw meaningful comparisons between different relabelling algorithms or to benchmark novel methods against the existing ones. Both issues are addressed to the \pkg{label.switching} package. At first, the output of each relabelling method is reported in a way that the resulting single best clusterings are comparable among them. Moreover, the user can provide alternative sets of permutations arising from any (consistent) relabelling procedure and directly compare them to the available methods.

The rest of the paper is organised as follows. A short introduction to mixture models is given at Section \ref{sec:notation}.  Section \ref{sec:labelswitching} discusses the label switching phenomenon and some helpful notation is introduced. The relabelling algorithms contained at the \pkg{label.switching} package are briefly reviewed at Section \ref{sec:algs}. Section \ref{sec:r.functions} gives an overview of the package and the most important functions are described in detail. The practical implementation of the package is illustrated at Section \ref{sec:apps} using two real datasets from classic mixture and hidden Markov models, as well as two simulated datasets from mixtures of bivariate normal distributions. The paper concludes at Section \ref{sec:discussion}.

\section{Mixture models}\label{sec:notation}

Let $\boldsymbol{x}=(x_1,\ldots,x_n)$ denote a sample of $n$ (possibly multivariate) observations. Assume that $\boldsymbol{z}=(z_1,\ldots,z_n)$ is an unobserved (latent) sequence of state variables, with $z_i\in\{1,\ldots,K\}$, where $K>1$ denotes a known integer. Let $f$ denotes a member of a parametric family of distributions $f\in\mathcal F_\Theta:=\{f(\cdot|\theta):\theta\in\Theta\}$, with $\Theta\subseteq\mathbb{R}^{d}$, for some $d\geqslant 1$. Conditionally to $z_i$ and a vector of parameters $\boldsymbol{\theta}=(\theta_1,\ldots,\theta_K)$ the observations are distributed according to
\begin{equation}
x_i|(z_i=k,\theta_k) \sim f(\cdot|\theta_k), \quad k=1,\ldots,K
\end{equation}
for $i=1,\ldots,n$. 

Assume that $z_i$, $i=1,\ldots,n$, are independent random variables following the multinomial distribution with weights $\boldsymbol w:=(w_1,\ldots,w_K)$, that is,
\begin{equation}\label{eq:z.ind}
P(z_i=k|w_k)=w_k, \quad k=1,\ldots,K
\end{equation}
independently for $i=1,\ldots,n$ and $\boldsymbol w \in \boldsymbol W:=\{w_k>0, k=1,\ldots,K-1: \sum_{k=1}^{K-1}w_k < 1; w_K:=1-\sum_{k=1}^{K-1}w_k\}$. The  marginal distribution of $x_i$ is a finite mixture of $K$ distributions: 
\begin{equation}\label{eq:mixture}
x_i|\boldsymbol{\theta},\boldsymbol{w}\sim\sum_{k=1}^{K}w_kf(x_i|\theta_k).
\end{equation}

The conditional probability for observation $i$ to belong to component $k$ can be expressed as
\begin{equation}\label{eq:probs}
p_{ik}=\frac{w_k f(x_i|\theta_k)}{w_1 f(x_i|\theta_1)+\ldots + w_K f(x_i|\theta_K)}, \quad i=1,\ldots,n;\quad k=1\ldots,K.
\end{equation} 
We will refer to Equation~\ref{eq:probs} with the term classification probabilities. The observed likelihood of the mixture is defined as:
\begin{equation}\label{eq:observed}
L(\boldsymbol\theta,\boldsymbol{w};\boldsymbol x)= \prod_{i=1}^{n}\sum_{k=1}^{K}w_{k}f(x_i|\theta_{k}).
\end{equation}
The complete likelihood of the model is written as:
\begin{equation}\label{eq:complete}
L_c(\boldsymbol\theta,\boldsymbol{w};\boldsymbol x, \boldsymbol z)= \prod_{i=1}^{n}w_{z_i}f(x_i|\theta_{z_i}).
\end{equation}

\subsection{Label switching phenomenon}\label{sec:labelswitching}

Let $\mathcal T_K$ denote the set of permutations of $\{1,\ldots,K\}$. For any  $\boldsymbol\tau=(\tau_1,\ldots,\tau_K)\in\mathcal T_K$ define the corresponding permutation of the component specific parameters, weights and allocations as: $\boldsymbol\tau\boldsymbol\theta : = (\theta_{\tau_1},\ldots,\theta_{\tau_K})$,  $\boldsymbol\tau\boldsymbol w : = (w_{\tau_1},\ldots,w_{\tau_K})$ and $\boldsymbol\tau\boldsymbol z:=(\tau_{z_1},\ldots,\tau_{z_n})$, respectively. 

Notice now that the likelihood of the mixture model is invariant for any permutation of the parameters, that is, 
\begin{equation}\label{eq:likinv}
L(\tau\boldsymbol{\theta},\tau\boldsymbol{w};\boldsymbol x)=L(\boldsymbol{\theta},\boldsymbol{w};\boldsymbol x),
\end{equation}
for all $\tau \in \mathcal T_K$, $\boldsymbol\theta\in\Theta$, $\boldsymbol w\in\boldsymbol W$. Let $p(\boldsymbol{\theta},\boldsymbol{w})$ denote the prior distribution of component specific parameters and weights of the model. It will be assumed that this prior is permutation invariant as well, that is, \begin{equation}\label{eq:priorinv}p(\tau\boldsymbol{\theta},\tau\boldsymbol{w})=p(\boldsymbol{\theta},\boldsymbol{w}),\end{equation} for all $\tau \in\mathcal T_K$, $\boldsymbol\theta\in\Theta$, $\boldsymbol w\in\boldsymbol W$. A typical choice  \citep{Richardson:97,Fruhwirth:01,Marin:05} for the prior of the component parameters is to assume that  $\boldsymbol{\theta}\sim\prod_{k=1}^{K}p(\theta_k)$, independently for $k=1,\ldots,K$, for a specific family of distributions $p(\cdot)$ which is common to all states. A common prior assumption on the mixture weights is a non-informative Dirichlet distribution.

Let  $p(\boldsymbol \theta, \boldsymbol w|\boldsymbol x)\propto L(\boldsymbol\theta,\boldsymbol{w};\boldsymbol x)p(\boldsymbol \theta,\boldsymbol w)$ denotes the posterior distribution of the mixture model parameters.  
From Equations~\ref{eq:likinv} and~\ref{eq:priorinv} it follows that the same invariance property holds for the posterior distribution, that is,  $p(\tau\boldsymbol{\theta},\tau\boldsymbol{w}|\boldsymbol x)=p(\boldsymbol{\theta},\boldsymbol{w}|\boldsymbol x)$, for all $\tau \in\mathcal T_K$, $\boldsymbol\theta\in\Theta$, $\boldsymbol w\in\boldsymbol W$. This implies that all marginal densities of component specific parameters and weights are coinciding.  Now, if a simulated output from any MCMC sampler has converged to the symmetric posterior distribution, the generated values will be switching between the symmetric high posterior density areas. 

This behaviour is known as the label switching phenomenon and makes the generated MCMC sample non-identifiable. Hence, it is not straightforward to draw inference for any parametric function that depends on the labels of the components. In order to derive meaningful estimates, all simulated parameters should be switched to one among the $K!$ symmetric areas of the posterior distribution. This is done by applying suitable permutations of the labels $\{1,\ldots,K\}$ to each MCMC draw.

\section{Algorithms}\label{sec:algs}

In this section we will describe the relabelling algorithms that are available to \pkg{label.switching} package. It consists of seven deterministic and one probabilistic relabelling method, as shown at Table \ref{tab:algorithms}. The third column describes the necessary input of the algorithms, while the input notation is described in detail at Table \ref{tab:notation}, where $m$ denotes the number of retained MCMC iterations. 

For practical purposes it will be convenient to arrange all component specific parameters ($\boldsymbol{\theta}$) and weights/transition probabilities ($\boldsymbol{w}$) at a $K\times J$ matrix $\boldsymbol\xi$. The number of columns ($J$) of the global parameter vector $\boldsymbol\xi:=(\boldsymbol{\theta},\boldsymbol{w})$ is equal to the number of different types of parameters of the model. For example, if Equation~\ref{eq:mixture} corresponds to a univariate normal mixture model, then there are $J=3$ different types: $\xi_{kj}$ denotes the mean ($j=1$), variance ($j=2$) and weight ($j=3$), respectively, for component $k = 1,\ldots,K$. In case of a bivariate normal mixture, there are $J=6$ parameter types for each component: two parameters for the mean vector, two variances, one covariance and one weight. 

In the sequel, we will assume that an augmented sample $(\boldsymbol{\xi}^{(t)},\boldsymbol{z}^{(t)})$, $t = 1,\ldots,m$, has been generated by an MCMC algorithm. Moreover, let $p_{ik}^{(t)}$, $t=1,\ldots,m$, $k = 1,\ldots,K$, $i=1,\ldots,n$ denote the corresponding classification probabilities across the MCMC run.

\begin{table}
\centering{
\begin{tabular}{l|l|l}
\hline
Function & Method & Input  \\
\hline
\code{aic} & Ordering constraints & \code{mcmc, constraint} \\
\code{dataBased} & Data-based & \code{z, x, K} \\
\code{ecr} & ECR (default) & \code{z, zpivot, K}\\
\code{ecr.iterative.1} & ECR (iterative vs.~1) & \code{z, K}\\
\code{ecr.iterative.2} & ECR (iterative vs.~2) & \code{z, K, p}\\
\code{pra} & PRA & \code{mcmc, pivot}\\
\code{stephens} & Stephens & \code{p}\\
\code{sjw} & Probabilistic & \code{mcmc, z, x, complete}\\
\hline
\end{tabular}
}
\caption{The available relabelling algorithms at \pkg{label.switching} package.}\label{tab:algorithms}
\end{table}

\begin{table}
\centering{
\begin{tabular}{l|l|l|l}
\hline
Object & Type & Dimension & Details\\
\hline
\code{z} & Array (integer) & $m\times n$& simulated allocation vectors\\
\code{x} & Array & $n\times d$& observed data\\
\code{zpivot} & Numeric (integer) & $n$ & pivot allocation vector\\
\code{p} & Array (real) & $m\times n \times K$& classification probabilities\\
\code{mcmc} & Array (real) & $m\times K\times J$& simulated parameters\\
\code{pivot} & Array (real) & $K\times J$& pivot parameter\\
\code{complete} & Function &  -- & complete log-likelihood function\\
\hline
\end{tabular}
}
\caption{Input notation for the relabelling algorithms.}\label{tab:notation}
\end{table}

\subsection{Ordering constraints}

Imposing an artificial identifiability constraint to the MCMC sample is the simplest approach to the label switching problem. In such a case, the simulated MCMC output is permuted according to the ordering of a specific parameter. However, this approach works well only in cases that the selected constraint is able to separate the symmetric posterior modes, which is rarely true.  

\begin{alg}[Ordering constraints] \hspace{1ex}\\
\begin{enumerate}
\item Choose a specific parameter type $\xi_s$, $s = 1,\ldots,J$.
\item For $t = 1,\ldots,m$ find the permutation $\tau^{(t)}\in\mathcal T_K$ that  $\xi_{\tau_1j}^{(t)}<\ldots<\xi_{\tau_Kj}^{(t)}$.
\end{enumerate}
\end{alg}

\subsection{Stephens' method}

One of the first principled solutions to the label switching problem was proposed by \citet{stephens}. The idea behind Stephens' algorithm is to make the permuted MCMC draws agree on the $n\times K$ matrix of classification probabilities. For this purpose, the Kullback-Leibler divergence between an averaged matrix of classification probabilities across the MCMC run and the classification matrix at each MCMC iteration is minimized at an iterative fashion. In general, Stephens' algorithm is very efficient in terms of finding the correct relabelling, but its drawback is the need to store the $m\times n \times K$ matrix \code{p} of classification probabilities. 

\begin{alg}[Kullback-Leibler relabelling]
\hspace{1ex}\\
\begin{enumerate}
\item Choose $m$ initial permutations $\tau^{(t)}$ $t=1,\ldots,m$ (usually set to identity).
\item For $t=1,\ldots,m$, $k=1,\ldots,K$ calculate $q_{ik}:=\frac{1}{m}\sum_{t=1}^{m}p_{i\tau_{k}}^{(t)}$.
\item For $t = 1,\ldots,m$ find a permutation $\tau^{(t)}\in\mathcal T_K$ that minimizes $\sum_{i=1}^{n}\sum_{k=1}^{K}p_{i\tau_k}^{(t)}\log\left(\frac{p_{i\tau_k}^{(t)}}{q_{ik}}\right)$.
\item If an improvement is made to $\sum_{t=1}^{m}\sum_{i=1}^{n}\sum_{k=1}^{K}p_{i\tau_k}^{(t)}\log\left(\frac{p_{i\tau_k}^{(t)}}{q_{ik}}\right)$ go to step 2, finish otherwise.
\end{enumerate}
\end{alg}

\subsection{Pivotal reordering algorithm}

The Pivotal Reordering Algorithm (PRA), proposed by \citet{Marin:05, Marin:07}, is a very simple geometrically-based solution to the label switching. The idea is to permute all simulated MCMC samples of parameters so that they are maximizing their similarity to a pivot parameter vector, as the complete MAP estimate. This is done by selecting the permutation that minimizes the Euclidean distance between the pivot and the set of permuted parameter vectors at each MCMC iteration. In principle, this method is a data-driven way to apply an artificial identifiability constraint on the parameter space.

\begin{alg}[Pivotal Reordering]
\hspace{1ex}\\
\begin{enumerate}
\item Define a pivot parameter vector: $\boldsymbol{\xi}^{*}=(\xi^{*}_{kj})$, $k=1,\ldots,K$, $j=1,\ldots,J$.
\item For $t = 1,\ldots,m$ find a permutation $\tau^{(t)}\in\mathcal T_K$ that maximizes $\sum_{j=1}^{J}\sum_{k=1}^{K}\xi^{(t)}_{\tau_k j}\xi_{kj}^{*}$.
\end{enumerate}
\end{alg}
Note here that maximizing the dot product $\tau\boldsymbol \xi^{(t)}\cdot\boldsymbol\xi^{*}$ in step 2 is equivalent to minimizing the Euclidean distance between $\tau\boldsymbol \xi^{(t)}$ and  $\boldsymbol\xi^{*}$.
\subsection{ECR algorithms}

ECR algorithm was originally proposed by \citet{Papastamoulis:10} and it is based on the idea that equivalent allocation vectors are mutually exclusive from the label switching solution. Two allocation vectors are called equivalent if the first one arises from the second by simply permuting its labels. ECR algorithm partitions the set of allocation vectors into equivalence classes and selects a representative from each class. Then, the permutation needed to be applied at a given MCMC iteration is determined by the one that reorders the corresponding allocations in order to become identical to the representative of its class.

In the default version of ECR algorithm (\code{ecr}), equivalence classes are determined using a pivot allocation vector \code{zpivot}. The pivot is selected by choosing a high-posterior density point, such as the complete or non-complete Maximum A Posteriori (MAP) estimate. \citet{Rodriguez} tried to relax the dependence of ECR algorithm to a pivot and proposed two iterative versions (\code{ecr.iterative.1} and \code{ecr.iterative.2}). The first algorithm is using as input only the simulated allocation variables and is initialized by a pivot selected at random. Then, the standard version of ECR is repeated until a fixed pivot has been found.  Nevertheless, it is not guaranteed that this procedure will lead to a ``good''  pivot. The second iterative ECR algorithm requires the knowledge of classification probabilities across the MCMC run and it could be described as an allocation vectors version of Stephens' algorithm. Of course the problem of storing the matrix \code{p} applies to this method as well. However, as it will be demonstrated in the applications, \code{ecr.iterative.2} is significantly faster than \code{stephens}.

\begin{alg}[ECR: default version]
\hspace{1ex}\\
\begin{enumerate}
\item Define a pivot allocation: $\boldsymbol{z}^{*}=(z^{*}_{1},\ldots,z^{*}_{n})$.
\item For $t = 1,\ldots,m$ find a permutation $\tau^{(t)}\in\mathcal T_K$ that maximizes $\sum_{i=1}^{n}I(\tau z_i^{(t)}=z^{*}_i)$.
\end{enumerate}
\end{alg}
\begin{alg}[ECR: iterative version 1]
\hspace{1ex}\\
\begin{enumerate}
\item Choose $m$ initial permutations $\tau^{(t)}$, $t=1,\ldots,m$ (usually set to identity).
\item Update the pivot: $z^{*}_i = \mbox{mode}\{\tau z_i^{(t)};t=1,\ldots,m\}$, $i=1,\ldots,n$.
\item For $t = 1,\ldots,m$ find a permutation $\tau^{(t)}\in\mathcal T_K$ that maximizes $\sum_{i=1}^{n}I(\tau z_i^{(t)}=z^{*}_i)$.
\item If an improvement is made to $\sum_{t=1}^{m}\sum_{i=1}^{n}I(\tau z_i^{(t)}=z^{*}_{i})$ go to step 2, finish otherwise.
\end{enumerate}
\end{alg}
\begin{alg}[ECR algorithm: iterative version 2]
\hspace{1ex}\\
\begin{enumerate}
\item Choose $m$ initial permutations $\tau^{(t)}$, $t=1,\ldots,m$ (usually set to identity).
\item Update the pivot: $z^{*}_i = \mbox{argmax}\{p_{i\tau_k}^{(t)};t=1,\ldots,m\}$, $i=1,\ldots,n$.
\item For $t = 1,\ldots,m$ find a permutation $\tau^{(t)}\in\mathcal T_K$ that maximizes $\sum_{i=1}^{n}I(\tau z_i^{(t)}=z^{*}_i)$.
\item If an improvement is made to $\sum_{t=1}^{m}\sum_{i=1}^{n}I(\tau z_i^{(t)}=z^{*}_{i})$ go to step 2, finish otherwise.
\end{enumerate}
\end{alg}

\subsection{Probabilistic relabelling algorithm}

Another method provided by the package is the probabilistic relabelling algorithm \code{sjw} of \citet{sperrin}. Under this concept, the permutation for each MCMC draw is treated as missing data with associated uncertainty. Then, an EM-type algorithm computes the expected values of $K!$ permutation probabilities per MCMC iteration, given an estimate of the parameter values. In the maximization step, this estimate is updated using a weighted average of all permuted parameters. This method requires a large amount of user input: the generated  MCMC sample of parameters and latent allocation variables, the observed data and a function that computes the complete log-likelihood. The algorithm is not efficient when the number of components grows large due to the computational overload.

\begin{alg}[Probabilistic relabelling]
\hspace{1ex}\\
\begin{enumerate}
\item Initialize an estimate of the parameters $\widehat{\boldsymbol{\xi}}=(\widehat{\boldsymbol{w}},\widehat{\boldsymbol{\theta}})$ and repeat steps 2 and 3 until a fixed point is reached.
\item E-Step: For $t=1,\ldots,m$, compute permutation probabilities
$g_{\tau}^{(t)}\propto L_c(\tau\widehat{\boldsymbol{\xi}}|\boldsymbol{z^{(t)}})$, $\tau\in\mathcal T_K$.
\item M-Step: Update parameter estimate $\widehat{\boldsymbol{\xi}}=\frac{1}{m}\sum_{t=1}^{m}\sum_{\tau\in\mathcal T_K}g_{\tau}^{(t)}\tau\boldsymbol{\xi}^{(t)}$.
\end{enumerate}
\end{alg}

\subsection{Data-based relabelling}

The data-based method of \citep{Rodriguez} is a deterministic relabelling algorithm. At first, a set of cluster centers $m_{kr}$ and dispersion parameters $s_{kr}$ is estimated for $k=1,\ldots,K$, $r=1,\ldots,d$. Next, the optimal permutations are defined as the ones minimizing a $k$-means type loss-function between the cluster pivots and the observed data, based on the simulated allocations at each MCMC iteration.   

\begin{alg}[Data-based relabelling]
\hspace{1ex}\\
\begin{enumerate}
\item Find estimates $m_{kr}$ and $s_{kr}$, $k=1,\ldots,K$, $r = 1,\ldots,d$.
\item For $t = 1,\ldots,m$, find a permutation $\tau\in\mathcal T_K$ that minimizes $$\sum_{k=1}^{K}\sum_{\ell=1}^{K}I(z_i^{(t)}=\tau_\ell)\sum_{\{i:\tau z_i^{(t)}=\ell\}}\sum_{r=1}^{d}\left(\frac{x_{ir}-m_{kr}}{s_{kr}}\right)^{2}.$$
\end{enumerate}
\end{alg}

The estimates at step 1 are solely based on the observed data $\boldsymbol x$ and the simulated allocation variables $\{\boldsymbol z^{(t)}$, $t=1,\ldots,m\}$. For more details, the reader is referred to algorithm 5 of \cite{Rodriguez}.

Finally, it is mentioned that algorithms \code{stephens}, \code{ecr}, \code{ecr.iterative.1}, \code{ecr.iterative.2} and \code{dataBased} are optimized using the library \pkg{lpSolve} \citep{lpSolve} for the solution of the assignment problem \citep{burkard}. This is a key-property for any computationally-efficient label switching solving algorithm, because in any other case the computational overload explodes as the number of components increases due to the computation of $K!$ quantities. By transporting the original problem into equivalent integer programming ones, the computational overload is avoided. The reader is referred to \citet{Rodriguez}. On the other hand, for each MCMC iteration, the \code{pra} and \code{sjw} algorithms require the computation of $K!$ dot products and permutation probabilities, respectively, so they are not suggested for large values of $K$.

\section{Implementation in R}\label{sec:r.functions}

All previously described relabelling algorithms are available as stand-alone functions at the \pkg{label.switching} package, as shown at Table \ref{tab:algorithms}. The input of each function is described at Table \ref{tab:notation}. Each one of them returns a list of permutations. The user can conveniently call any combination of these methods using the function \code{label.switching}, which serves as the main call function of the package. Moreover, a set of user-defined permutations can be also supplied which is useful for comparison purposes. In this section we will describe the general call of \code{label.switching} and explain the input arguments and output values in detail. 

\subsection{Structure of main function}

The general usage is
\begin{Sinput}
R> label.switching(method, zpivot, z, K, prapivot, p, complete, mcmc, 
+    sjwinit, data, constraint, groundTruth, thrECR, thrSTE, thrSJW,
+    maxECR, maxSTE, maxSJW, userPerm)
\end{Sinput}
and the details of the implementation are described in the sequel. 
\begin{itemize}[leftmargin=1.1in]
\item[\code{method}] the desired combination of the available methods. It can be any non-empty subset of:\\
\code{c("ECR", "ECR-ITERATIVE-1", "ECR-ITERATIVE-2", "PRA",\\"STEPHENS", "SJW", "AIC", "DATA-BASED")}\\
Also available is the option \code{"USER-PERM"} which corresponds to a user-defined set of permutations \code{userPerm}.
\item[\code{zpivot}] Obligatory only when \code{"ECR"} has been selected. It is a user-specified set of $d\geqslant 1$ pivots and it should be defined as an $d\times n\times K$ array. Each pivot should correspond to a high posterior-density area. Then, method \code{"ECR"} will be applied $d$ times.
\item[\code{z}]  $m\times n$-dimensional array corresponding to the set of simulated allocation vectors $\boldsymbol z^{(t)}$, $t=1,\ldots,m$, with $z_i^{(t)}\in\{1,\ldots,K\}$, for all $t = 1,\ldots,K$. It is required by: \code{"ECR"}, \code{"ECR-ITERATIVE-1"}, \code{"ECR-ITERATIVE-2"}, \code{"SJW"} and \code{"DATA-BASED"}.
\item[\code{K}] Positive integer (at least equal to 2) indicating the number of mixture components. It is required by: \code{"ECR"}, \code{"ECR-ITERATIVE-1"} and \code{"DATA-BASED"}. If missing, then it is set to $\max\{z_i^{(t)}:t=1,\ldots,m;i=1,\ldots,n\}$.
\item[\code{prapivot}] Obligatory only when \code{"PRA"} has been selected. It is a user-specified $K\times J$ array corresponding to a high posterior-density area for the parameters of the mixture.
\item[\code{p}] $m\times n\times K$ matrix of classification probabilities as defined in Equation~\ref{eq:probs}. Required by methods \code{"STEPHENS"} and \code{"ECR-ITERATIVE-2"}.
\item[\code{complete}] Complete log-likelihood function of the model. Required by method \code{SJW}. The input should be a $K\times J$ vector of parameters as well as an $n$-dimensional vector of allocations. The function should return a single value which corresponds to the complete log-likelihood as defined in Equation~\ref{eq:complete}. 
\item[\code{mcmc}] $m\times K\times J$ array of simulated parameters across the MCMC run. Required by methods \code{"PRA"}, \code{"SJW"} and \code{"AIC"}. 
\item[\code{sjwinit}] An index on the set $\{1,\ldots,m\}$ pointing at the MCMC iteration whose parameters will initialize the \code{sjw} algorithm (optional).
\item[\code{data}] the observed data $\boldsymbol x=(x_1,\ldots,x_n)$. Required by \code{"SJW"} and \code{"DATA-BASED"} methods.
\item[\code{constraint}] An (optional) integer between 1 and $J$ corresponding to the parameter that will be used to apply the Ordering Constraint.  If \code{constraint = "ALL"}, all $J$ ordering constraints are applied. Default value: 1.
\item[\code{groundTruth}] Optional integer vector of $n$ allocations, which are considered as the ``true'' allocations of the observations. The output of all methods will be relabelled in a way that the resulting single best clusterings maximize their similarity with the ground truth.
\item[\code{thrECR}] An (optional) positive threshold controlling the convergence criterion for \code{ecr.iterative.1} and \code{ecr.iterative.2}. Default value: $10^{-6}$.
\item[\code{thrSTE}] An (optional) positive threshold controlling the convergence criterion for \code{stephens}. Default value: $10^{-6}$.
\item[\code{thrSJW}] An (optional) positive threshold controlling the convergence criterion for \code{sjw}. Default value: $10^{-6}$.
\item[\code{maxECR}] An (optional) integer controlling the maximum number of iterations for \code{ecr.iterative.1} and \code{ecr.iterative.2}. Default value: 100.
\item[\code{maxSTE}] An (optional) integer controlling the maximum number of iterations for \code{stephens}. Default value: 100.
\item[\code{maxSJW}] An (optional) integer controlling the maximum number of iterations for \code{sjw}. Default value: 100.
\item[\code{userPerm}]  An (optional) list with $S$ user-defined permutations ($S\geqslant 1$). It is required only if \code{"USER-PERM"} has been chosen in \code{method}. In this case, \code{userPerm[[i]]} is an $m\times K$ array of permutations, $i = 1,\ldots,S$.
\end{itemize}

Let $f$ denotes the number of selected relabelling algorithms. The following values are returned.

\begin{itemize}[leftmargin=1.1in]
\item[\code{permutations}]  A list of $f$ permutation arrays: \code{permutations[[i]][t,]} corresponds to the permutation that must be applied to the parameters generated at the $t$-th MCMC iteration, according to method $i$, $i=1,\ldots,f$; $t=1,\ldots,m$.  
\item[\code{clusters}] $n$-dimensional vector of best clustering of the observations for each method.
\item[\code{timings}] the CPU time for the reordering part of each method, that is, the time to find the optimal permutations without taking into account the time spent by the user in order to compute the necessary input.
\item[\code{similarity}] $f'\times f'$ similarity matrix between the label switching solving methods in terms of their matching best-clustering allocations, where $f' = f$ if \code{groundTruth} is not supplied and $f' = f+1$ in the opposite case.
\end{itemize}

The output of the \code{label.switching} function is reported in a way that all relabelling methods maximize the similarity of the estimated single best clusterings with respect to a reference allocation vector. For this purpose, the number of matching allocations between two vectors is used. This makes easier the comparison between the different methods. By default, the reference allocation vector corresponds to the estimated single best clustering according to the first algorithm provided in \code{method}. In case that \code{groundTruth} is supplied by the user,  the reference allocation is set to the true one which is quite helpful in simulation studies.  

It is evident that each algorithm requires different types of input. Methods \code{aic}, \code{dataBased} and \code{ecr-iterative-1} require only quantities that are directly available from the raw MCMC output and/or the observed data. The algorithms \code{ecr}, \code{ecr-iterative-2}, \code{pra} and \code{stephens} demand a few extra lines of coding that mainly handle quantities that are already in use while the MCMC sampler is running. Finally, \code{sjw} is more demanding as the user has to provide a function along with the MCMC output.

The supplementary function \code{permute.mcmc}  reorders the MCMC sample (as stored in \code{mcmc}) according to the permutations returned by \code{label.switching}.

Usage: \begin{Sinput}
R> permute.mcmc(mcmc, permutations)
\end{Sinput}
Arguments:
\begin{itemize}[leftmargin=1.1in]
\item[\code{mcmc}] $m\times K\times J$ array containing an MCMC sample.
\item[\code{permutations}] $m\times K$ array of  permutations.
\end{itemize}
Value: 
\begin{itemize}[leftmargin=1.1in]
\item[\code{output}] reordered \code{mcmc} according to \code{permutations}.
\end{itemize}

\section{Examples}\label{sec:apps}

\subsection{Mixture of normal distributions: fishery data}\label{sec:mix}

\begin{figure}
\begin{tabular}{cc}
\includegraphics[width=.45\textwidth]{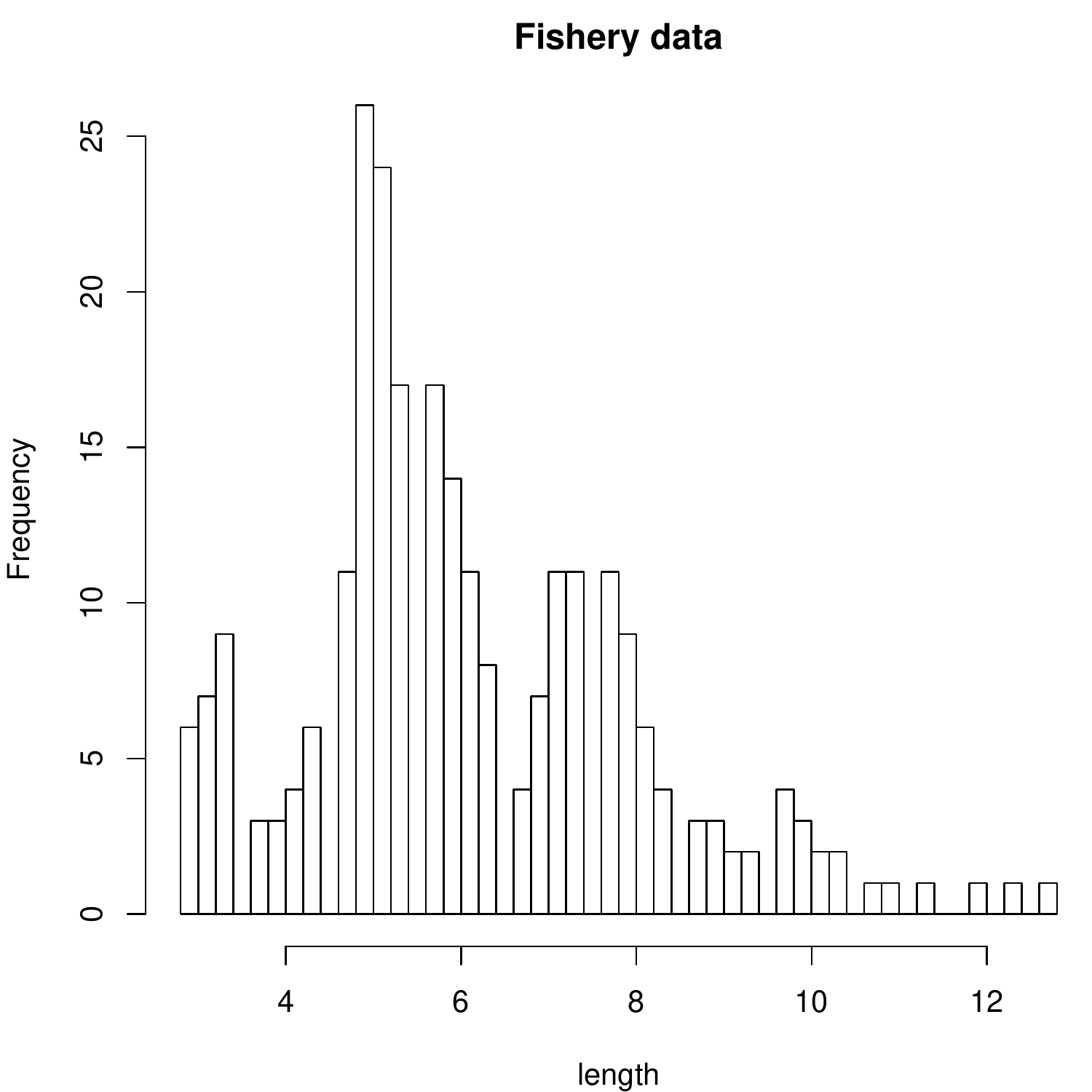}&
\includegraphics[width=.45\textwidth]{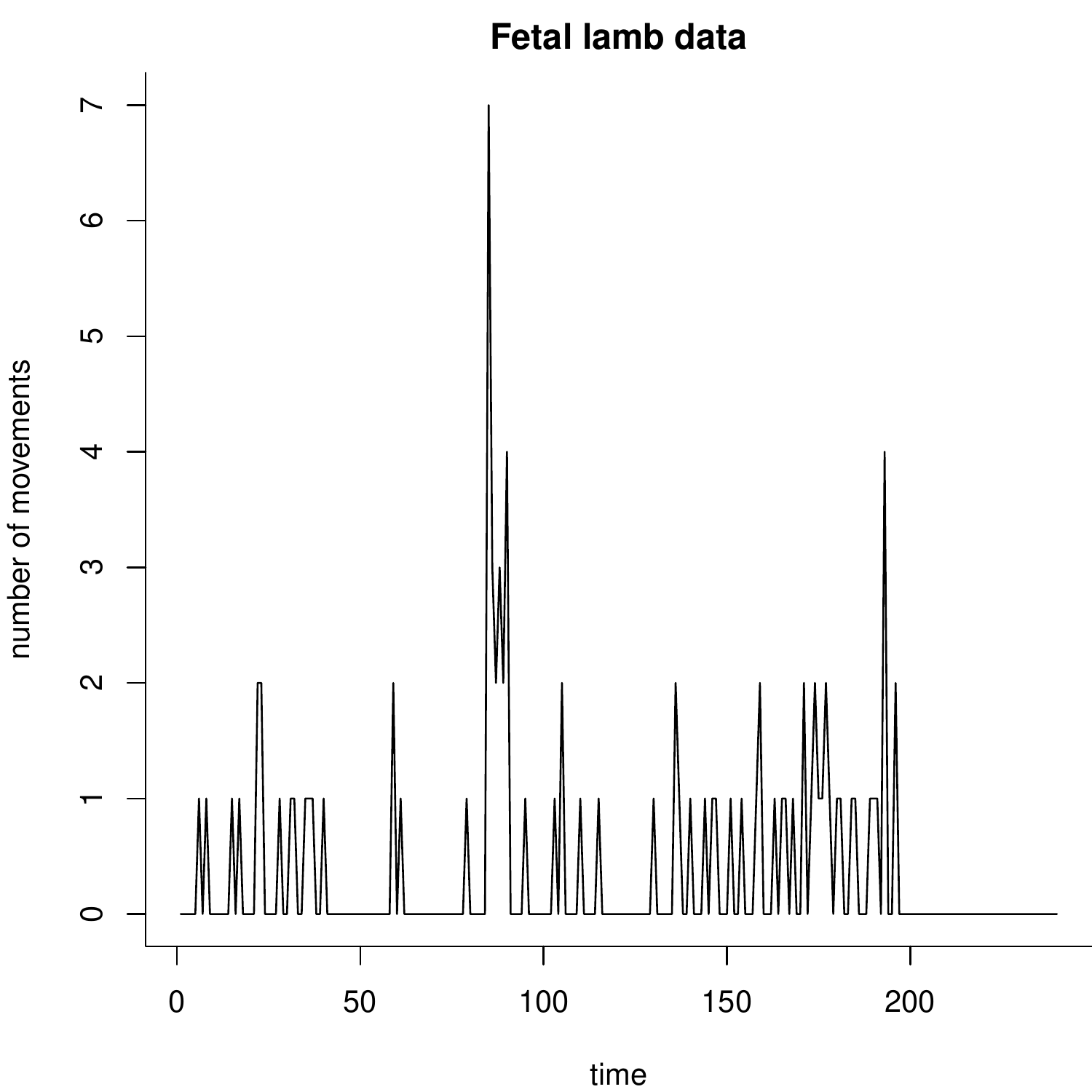}
\end{tabular}
\caption{Histogram of the fishery data and time series of the fetal lamb movements.}\label{fig:hist}
\end{figure}

The fishery data is taken from \citet{fish} and it consists of $n=256$ snapper length measurements. The histogram of the data is shown in Figure \ref{fig:hist} (left) and it is obvious that the length of a randomly sampled fish exhibits strong heterogeneity. This is due to the fact that the age of each fish has not been recorded. The data has been previously analysed as a mixture of $K$ normal distributions, that is,
$$x_i\sim \sum_{k=1}^{K}w_k\mathcal N(\mu_k,\sigma^{2}_k),$$
independent for $i=1,\ldots,n$. According to  \citet{Fruhwirth:06}, the number of components ranges from 3 to 5 and there are clearly four separated clusters in the MCMC draws. Here, we will consider a more challenging scenario with $K=5$ components. The MCMC sampler described in package \pkg{bayesmix} \citep{bayesmix} is applied in order to simulate an MCMC sample of $m=10000$ iterations from the posterior distribution of $\{\boldsymbol{z},\boldsymbol{\mu},\boldsymbol{\sigma^2},\boldsymbol{w}\}$, following a burn-in period of $1000$. This is done with the following commands.

\begin{Sinput}
R> library("bayesmix")
R> data("fish", package = "bayesmix")
R> x <- fish[ , 1]
R> n <- length(x)
R> K <- 5
R> m <- 11000 
R> burn <- 1000
R> model <- BMMmodel(fish, k = K, initialValues = list(S0 = 2),
+    priors = list(kind = "independence", parameter = "priorsFish",
+    hierarchical = "tau"))
R> control <- JAGScontrol(variables = c("mu", "tau", "eta", "S"),
+    burn.in = burn, n.iter = m, seed = 10)
R> mcmc <- JAGSrun(fish, model = model, control = control)
\end{Sinput}

\begin{figure}
\begin{tabular}{ccc}
\includegraphics[width=.3\textwidth]{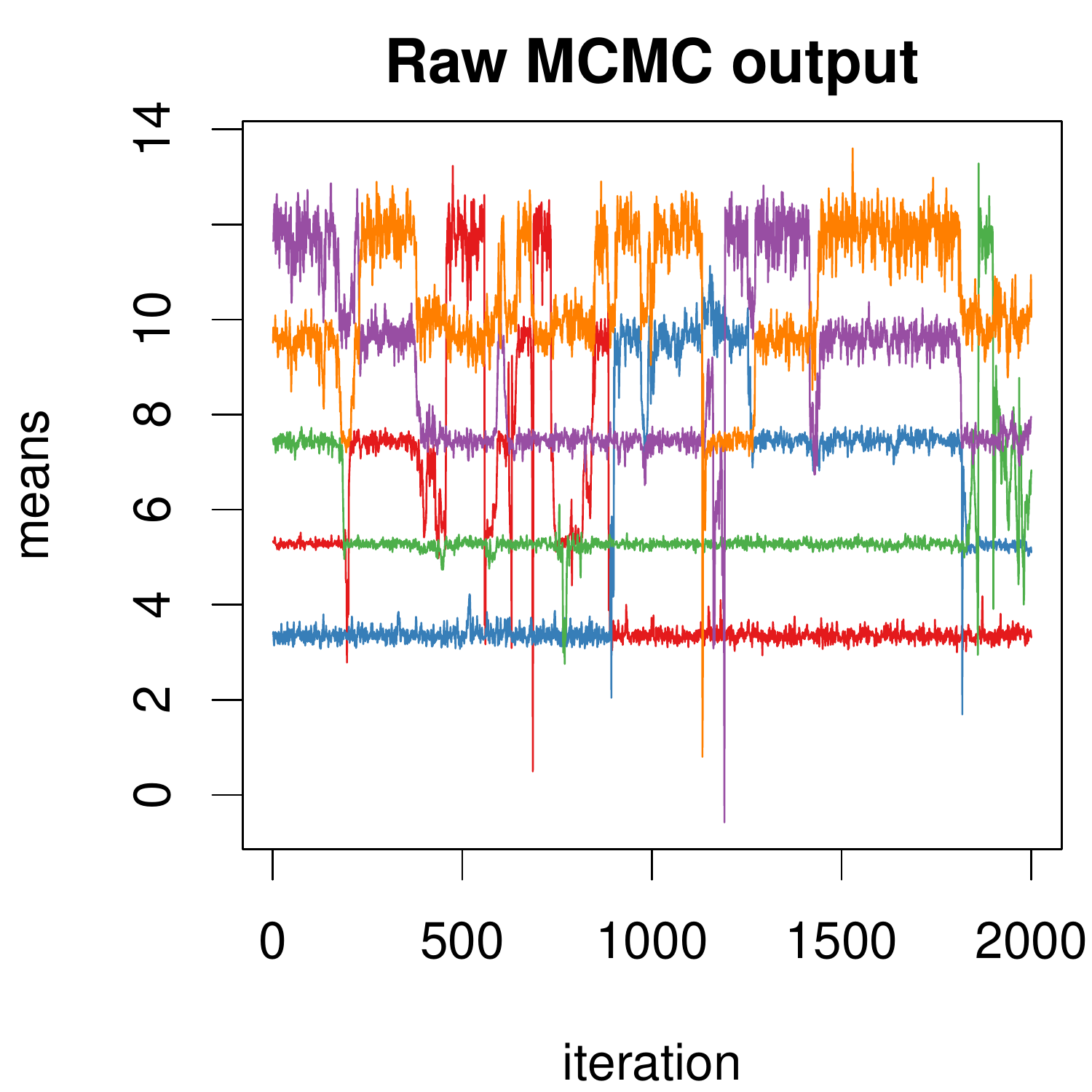}&
\includegraphics[width=.3\textwidth]{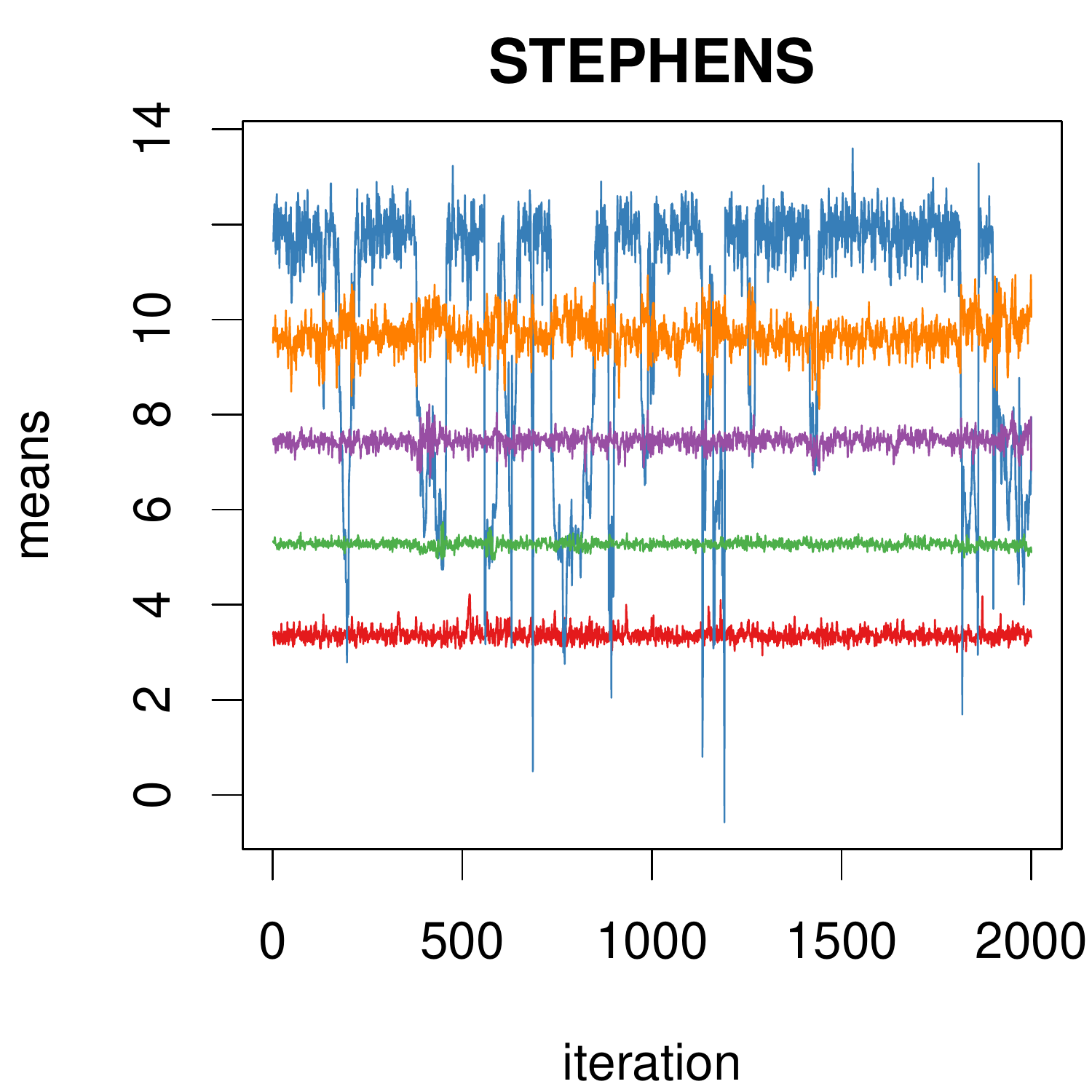}&
\includegraphics[width=.3\textwidth]{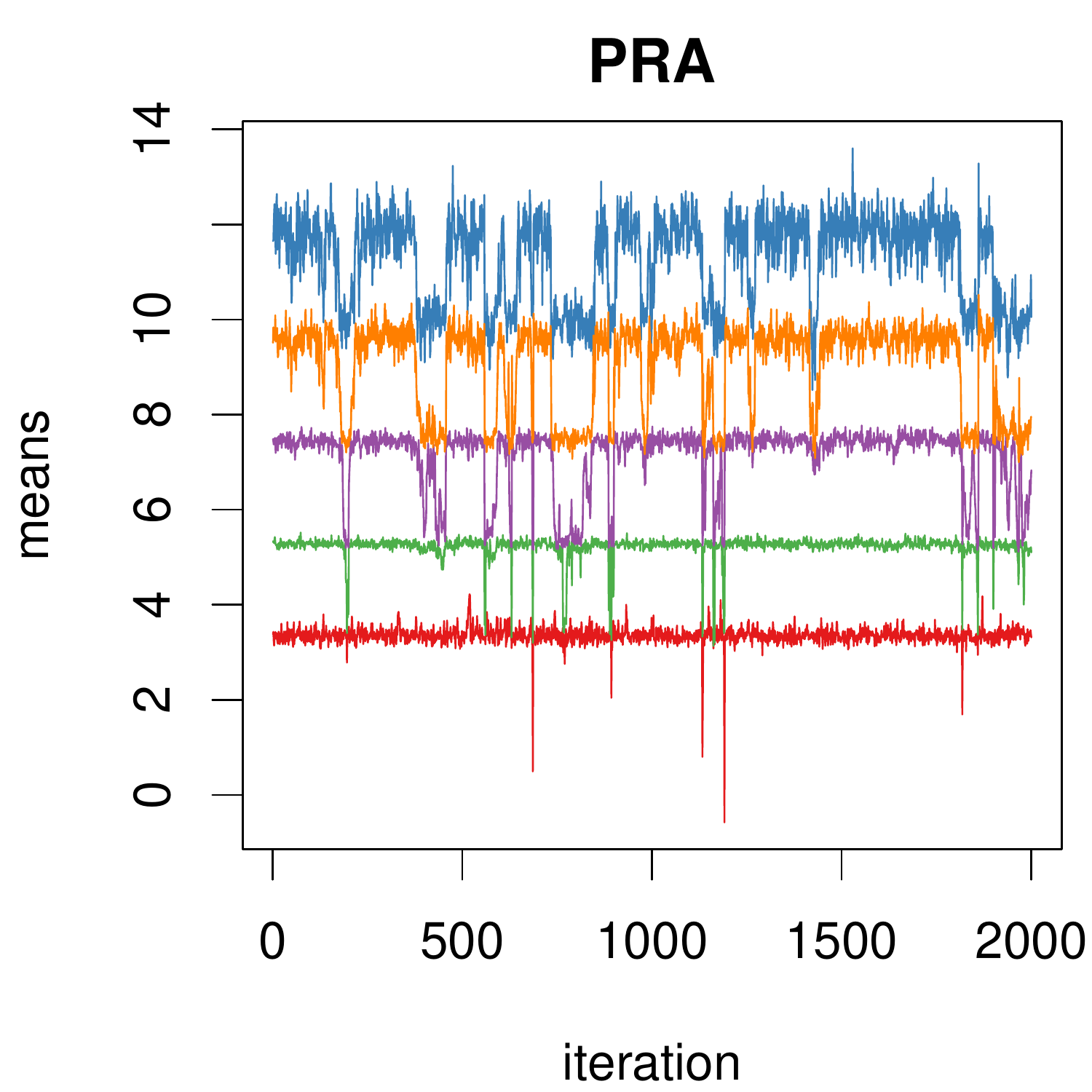}\\
\hspace{6.5ex}(a)&\hspace{6.5ex}(b)&\hspace{6.5ex}(c)\\
\includegraphics[width=.3\textwidth]{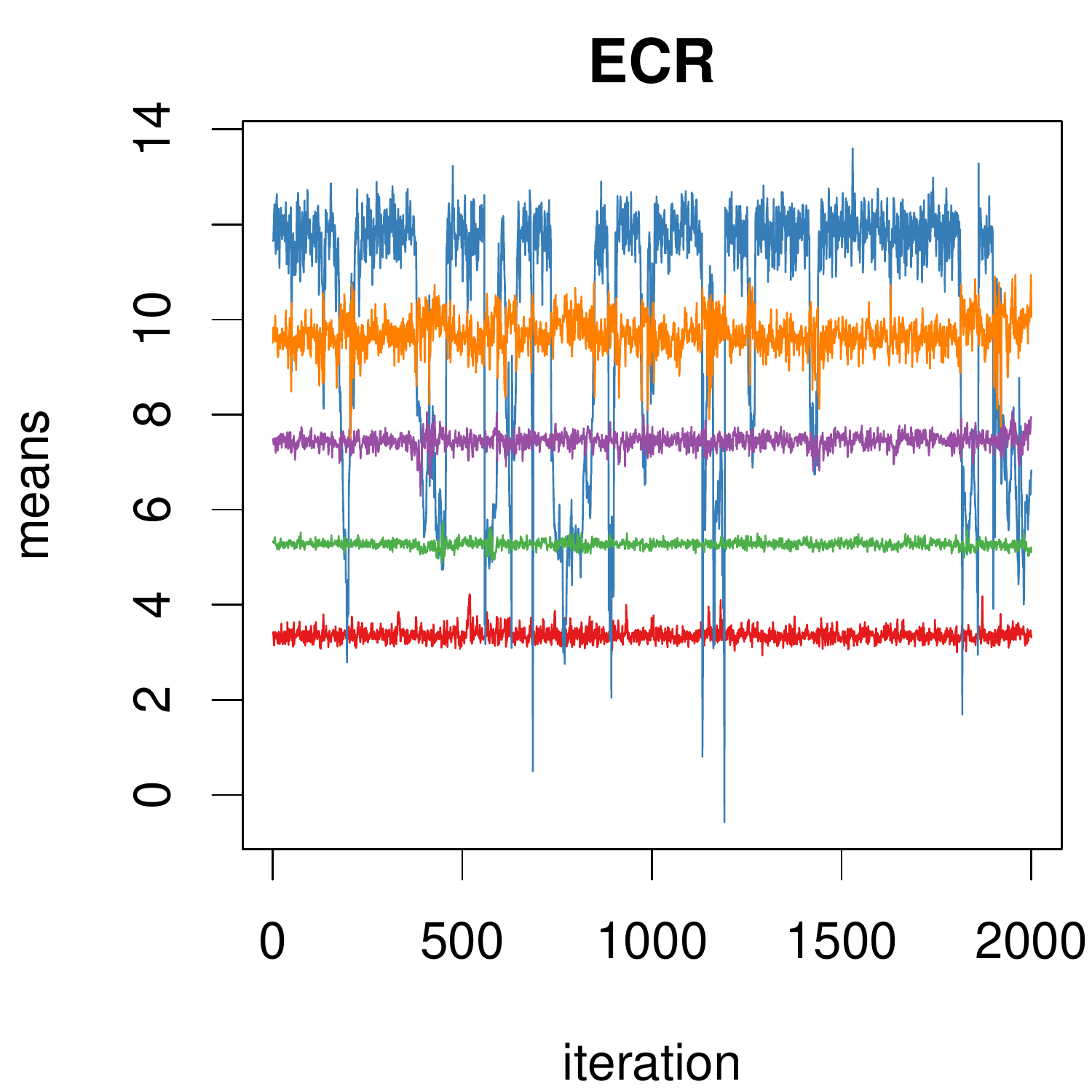}&
\includegraphics[width=.3\textwidth]{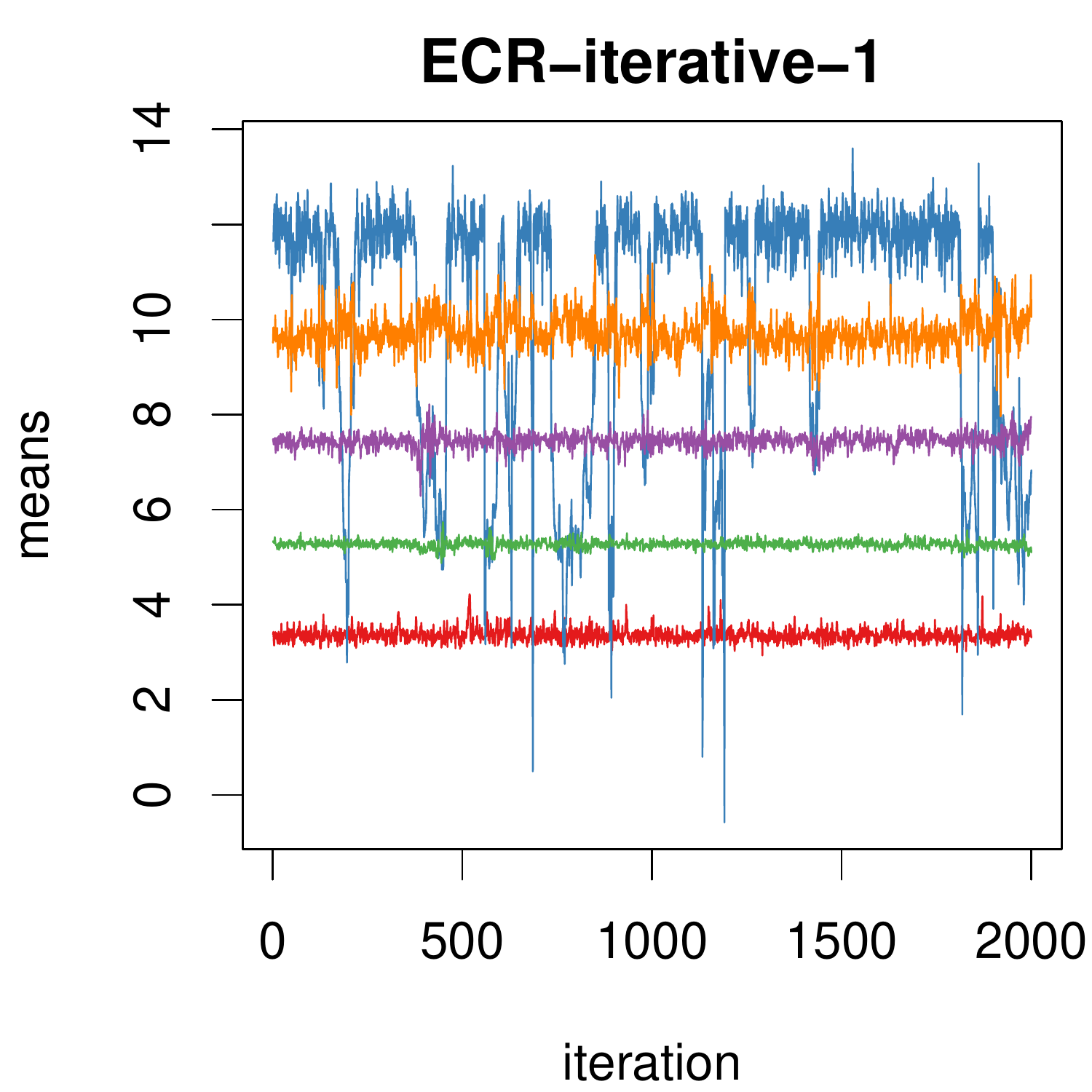}&
\includegraphics[width=.3\textwidth]{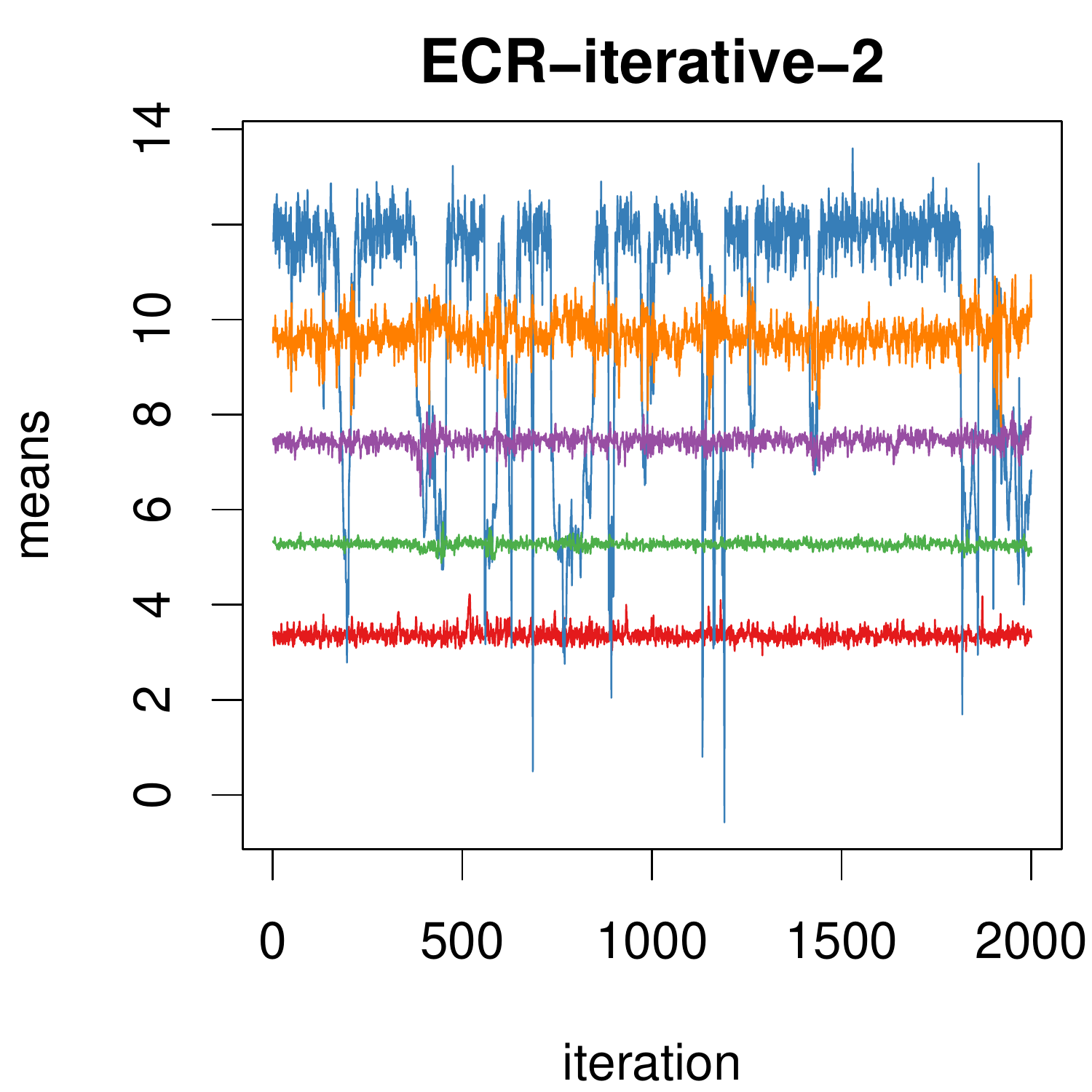}\\
\hspace{6.5ex}(d)&\hspace{6.5ex}(e)&\hspace{6.5ex}(f)\\
\includegraphics[width=.3\textwidth]{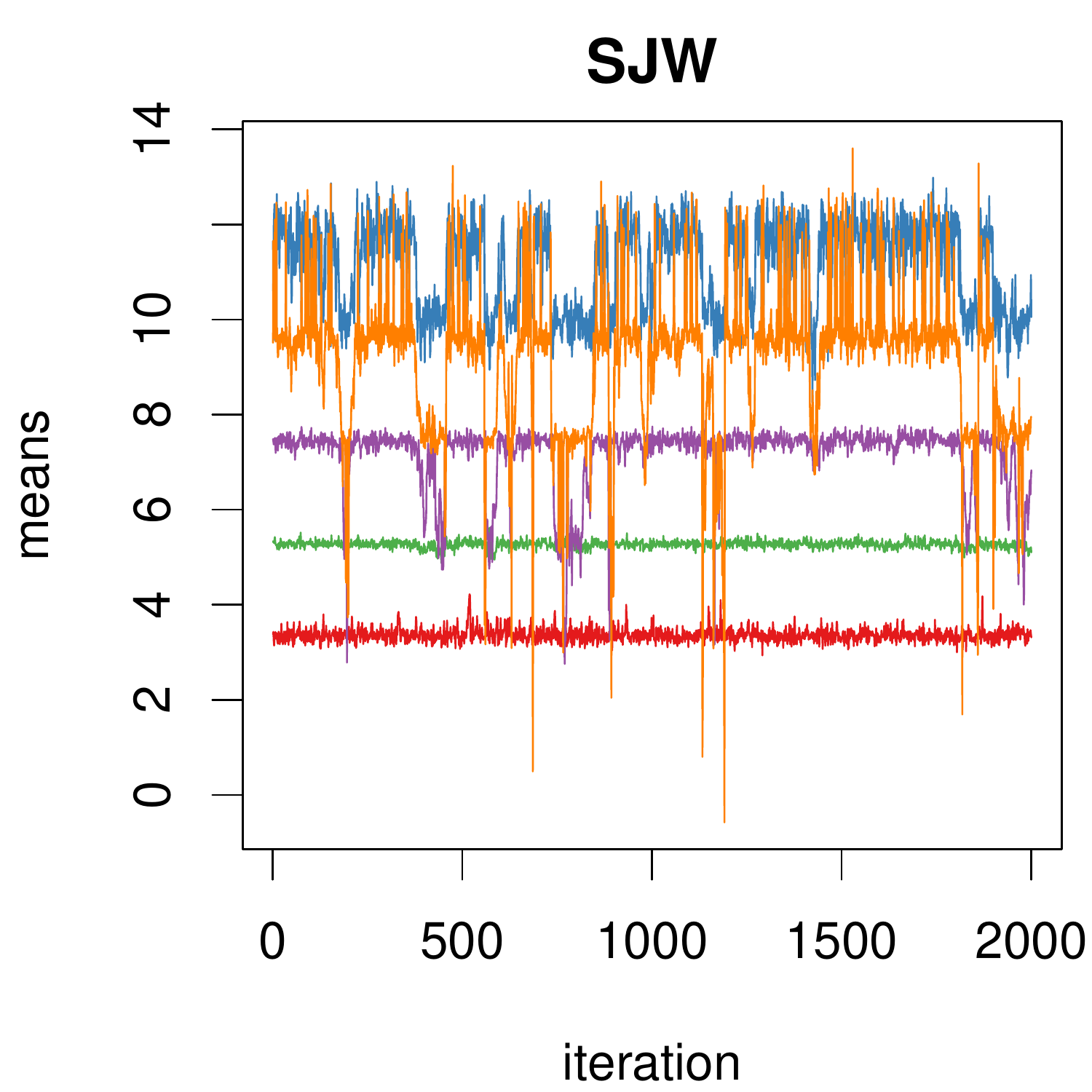}&
\includegraphics[width=.3\textwidth]{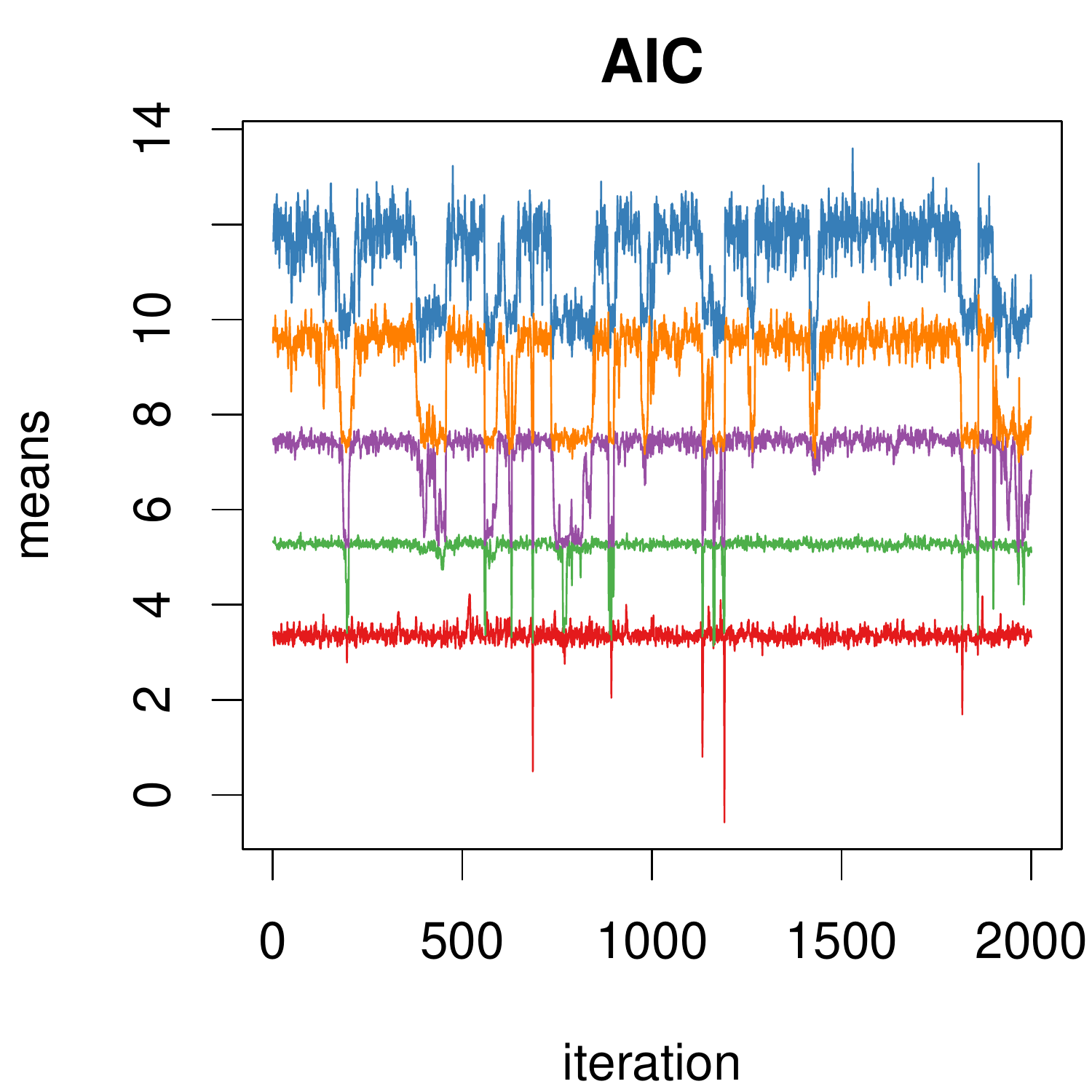}&
\includegraphics[width=.3\textwidth]{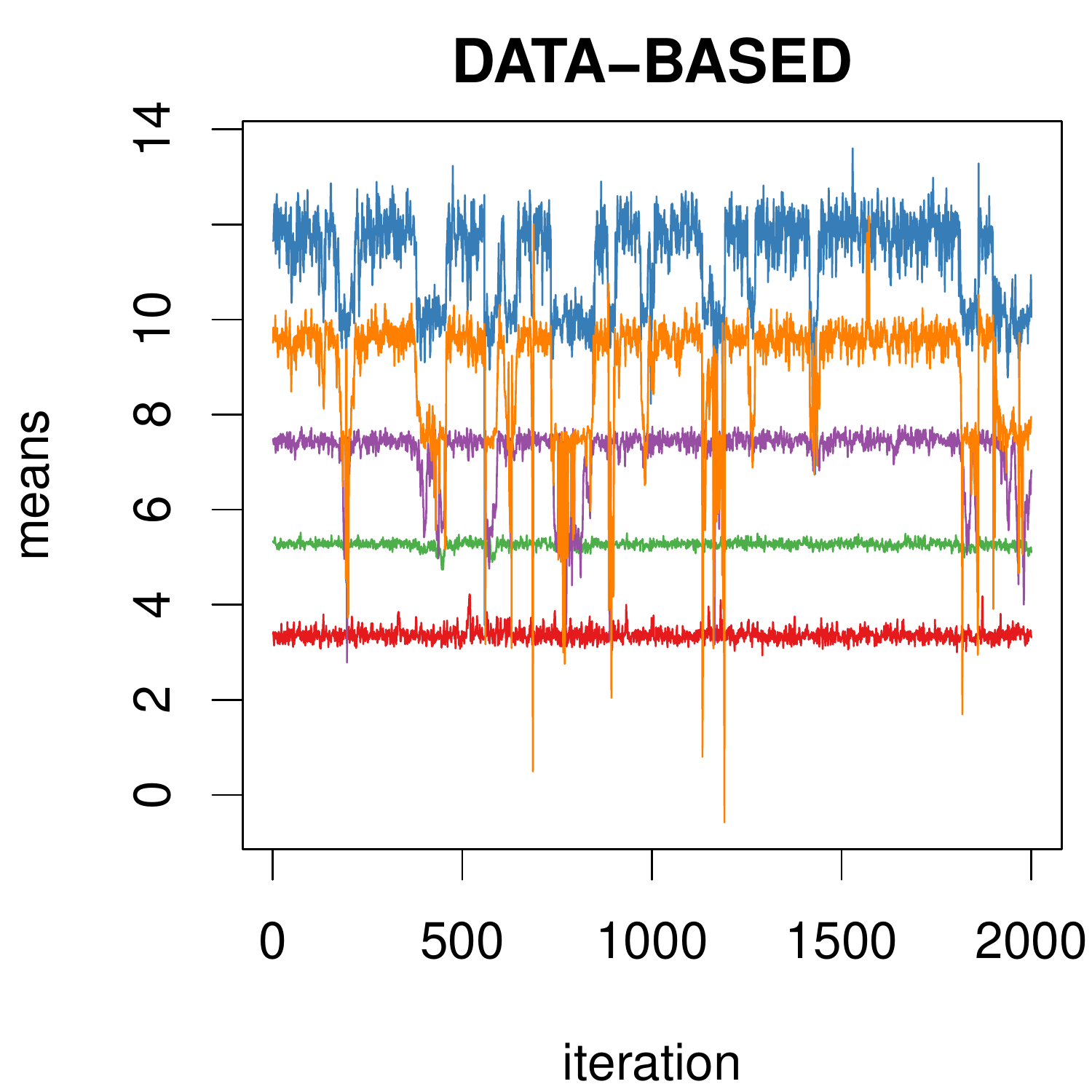}\\
\hspace{6.5ex}(g)&\hspace{6.5ex}(h)&\hspace{6.5ex}(i)
\end{tabular}
\caption{Fishery data. (a): Raw MCMC sample for $\mu_{k}$. (b), (c), (d), (e), (f), (g), (h), (i): Reordered MCMC sample by applying the permutations returned by \code{label.switching} function, according to methods: Stephens, PRA, ECR, ECR-iterative-1, ECR-iterative-2, SJW, AIC and Data-based, respectively.}\label{fig:fishery}
\end{figure}

The raw MCMC output for $\mu_{k}$, $k=1, \ldots,5$ is shown at Figure \ref{fig:fishery}.(a) (every 5th iteration displayed). It is obvious that the label switching phenomenon has occurred. Note that a simple ordering constraint to the means is not able to successfully isolate one of the symmetric high posterior-density areas. Next, we will apply the function \code{label.switching} considering all the presented relabelling algorithms. In order to do this, we have to compute all the related information that is required as input for each method. At first,  the MCMC output is converted into an $m\times K\times J$ array (\code{mcmc.pars}), where $J=3$ denotes the number of different parameter types for the normal mixture model:   means (\code{mcmc.pars[ , ,1]}), variances (\code{mcmc.pars[ , , 2]}) and weights (\code{mcmc.pars[ , ,3]}). Finally, the generated allocation variables are stored to $m\times n$ array \code{z}:
\begin{Sinput}
R> J <- 3
R> mcmc.pars <- array(data = NA, dim = c(m, K, J))
R> mcmc.pars[ , , 1] <- mcmc$results[-(1:burn), (n+K+1):(n+2*K)]	
R> mcmc.pars[ , , 2] <- mcmc$results[-(1:burn), (n+2*K+1):(n+3*K)]	
R> mcmc.pars[ , , 3] <- mcmc$results[-(1:burn), (n+1):(n+K)]	
R> z <- mcmc$results[-(1:burn), 1:n]
\end{Sinput}

Stephens' method as well as the second iterative version of ECR algorithm need the $m\times n\times K$ array of component membership probabilities $p_{ik}$, as defined in Equation~\ref{eq:probs}, for each MCMC iteration. These probabilities are stored to array \code{p} as follows.

\begin{Sinput}
R> p <- array(data = NA, dim = c(m, n, K))
R> for (iter in 1:m){
+    for(i in 1:n){
+    kdist <- mcmc.pars[iter, , 3]*dnorm(x[i], mcmc.pars[iter, , 1],
+    sqrt(mcmc.pars[iter, , 2]))
+    skdist <- sum(kdist)
+    for(j in 1:K){
+    p[iter, i, j] = kdist[j]/skdist}}}
\end{Sinput}

Method \code{sjw} demands to provide as input a function that computes the complete log-likelihood of the classic mixture model, as defined by taking the logarithm of Equation~\ref{eq:complete}. The next code accepts as input a dataset of univariate observations (\code{x}), an $n$-dimensional integer vector of allocations (\code{z}) and a $K\times J$ array of mixture parameters (means, variances, weights). 

\begin{Sinput}
R> complete.normal.loglikelihood <- function(x, z, pars){
+    g <- dim(pars)[1]
+    n <- length(x)
+    logl <- rep(0, n)
+    logpi <- log(pars[ , 3])
+    mean <- pars[ , 1]
+    sigma <- sqrt(pars[ , 2])
+    logl <- logpi[z] + dnorm(x, mean = mean[z], sd = sigma[z], log = T)
+    return(sum(logl))}
\end{Sinput}

The function \code{complete.normal.loglikelihood} will be also used for the determination of an MCMC iteration that corresponds to a high density area. Next, the allocation and parameters of this iteration will be used as pivot by the functions \code{ecr} and \code{pra}, respectively. After evaluating the complete log-likelihood function for the 10000 MCMC iterations, we obtained that the maximum value corresponds to iteration \code{mapindex = 4839}. 

We will also use an ordering constraint to the simulated means. Since this parameter type corresponds to \code{mcmc.pars[ , , j]} for \code{j = 1}, we should use \code{constraint = 1}. Now, we can apply  the available algorithms using the following command.
\begin{Sinput}
R> library("label.switching")
R> set <- c("STEPHENS", "PRA", "ECR", "ECR-ITERATIVE-1", "ECR-ITERATIVE-2",
+    "SJW", "AIC", "DATA-BASED")
R> ls <- label.switching(method = set, zpivot = z[mapindex, ], z = z,  K = K,
+    prapivot = mcmc.pars[mapindex, , ], p = p, constraint = 1,
+    sjwinit = mapindex, complete = complete.normal.loglikelihood,
+    mcmc = mcmc.pars, data = x)
R> ls$timings  
\end{Sinput}

\begin{table}[t]
\centering
\begin{tabular}{rrrr|rrrrrrrr}
  \hline
Data & $K$& $n$ & $m$ &steph & pra & ecr& ecr-1 & ecr-2 & sjw & aic & d-based \\ 
  \hline
Fish &$5$ & $256$ & $10000$ &56.9 & 3.5 & 3.4 & 17.8 & 13.4 & 613.7 & 0.1 & 11.4 \\ 
Lamb & $4$ &$240$ & $10000$ &61.0 & 0.8 & 2.7 & 11.4 & 11.0 &  598.5& 0.1 & 8.5 \\ 
Multivariate 1&$4$& $100$& $10000$ & 25.7 & 0.8 & 2.5& 10.7 & 6.0 & 695.9 & 0.1 & 11.0 \\
Multivariate 2&$9$ &$280$& $15000$ & 431.3 & NA & 9.1& 58.1 & 50.2 & NA & 0.1 & 61.1 \\
   \hline
\end{tabular}\caption{CPU times in seconds per relabelling method.}\label{tab:fish-times}
\end{table}

The last command returns the CPU time per method, which is shown at first line of Table \ref{tab:fish-times}. The MCMC draws can be reordered applying the permutations contained in \code{ls\$permutations} using the function \code{permute.mcmc}. Figure \ref{fig:fishery}(b)-(h) contain the reordered output for the simulated values of $\mu_{k}$, $k=1,\ldots,5$ (every 5-th iteration is displayed). We conclude that the results of methods \code{ecr}, \code{ecr-iterative-1}, \code{ecr-iterative-2} and \code{stephens} are quite similar to each other. 
The reordered values indicate that the component with the largest mean (blue coloured trace) exhibits a multimodal behaviour: there is a main mode at 12 and a minor one between 5 and 6.  On the other hand, \code{pra} and \code{dataBased} algorithms are driven by the generated values of the means and the resulting reordering is quite similar to the one that results from an ordering constraint to $\mu_k$. A similar behaviour is observed from \code{sjw} algorithm.  

Finally, the function \code{label.switching} returns the single best clusterings of the $n$ observations among the $K$ groups. This is simply done by calculating the mode of the reordered allocation vectors $z_i$, $i=1, \ldots,n$. The proportion of the matching allocations between any pair of the available methods is returned by \code{ls\$similarity} and it is shown at the lower diagonal of Table \ref{tab:fish-sim}.

\begin{table}[t]
\centering
\begin{tabular}{rrrrrrrrr}
  \hline
              & \code{steph} & \code{pra} & \code{ecr} & \code{ecr-1} & \code{ecr-2} & \code{sjw} &\code{aic} &\code{d-based} \\ 
  \hline
  \code{stephens} &       & 0.950 & 0.992 & 0.996 & 0.983 & 0.950 & 0.879 & 0.842 \\ 
       \code{pra} & 0.996 &       & 0.958 & 0.954 & 0.938 & 0.904 & 0.833 & 0.863 \\ 
       \code{ecr} & 1.000 & 0.996 &       & 0.996 & 0.975 & 0.942 & 0.871 & 0.850 \\ 
\code{ecr-iter-1} & 1.000 & 0.996 & 1.000 &       & 0.979 & 0.946 & 0.875 & 0.846 \\ 
\code{ecr-iter-2} & 1.000 & 0.996 & 1.000 & 1.000 &       & 0.966 & 0.899 & 0.825\\ 
       \code{sjw} & 0.992 & 0.996 & 0.992 & 0.992 & 0.992 &       & 0.896 & 0.792 \\ 
       \code{aic} & 0.996 & 1.000 & 0.996 & 0.996 & 0.996 & 0.996 &       & 0.720\\
   \code{d-based} & 0.996 & 1.000 & 0.996 & 0.996 & 0.996 & 0.996 & 1.000 & \\
   \hline
\end{tabular}
\caption{ \code{ls\$similarity}: Proportion of matching allocations for the single best-clusterings for each relabelling algorithm: Lower diagonal: Fish data. Upper diagonal: Fetal lamb data.}\label{tab:fish-sim}
\end{table}

\subsection{Poisson hidden Markov model: fetal lamb data}\label{sec:hmm}

A generalization of the classic mixture model set-up is to assume that the latent variables are forming an (unobserved) Markov chain. Let $\boldsymbol w=(w_{\ell k})$, $\ell,k \in \{1,\ldots,K\}$, denote an $K\times K$ matrix of transition probabilities. The conditional distribution of latent variables now is written as:
\begin{equation}\label{eq:z.markov}
P(z_i=k|z_{i-1}=\ell,\boldsymbol{w})=w_{\ell k}, \quad k=1,\ldots,K.
\end{equation}
The sequence $(z_1,\ldots,z_n)$ is unobserved and this justifies the term hidden Markov model. In this case, the complete likelihood is defined as:
\begin{equation}\label{eq:complete2}
L_c(\boldsymbol\theta,\boldsymbol{w}|\boldsymbol x, \boldsymbol z)= \pi_{z_1}f(x_1|\theta_{z_1})\prod_{i=2}^{n}w_{z_{i-1}z_{i}}f(x_i|\theta_{z_{i}}),
\end{equation}
where $\pi_k$, $k=1, \ldots,K$ denotes the left eigenvector of the transpose transition matrix $\boldsymbol{w}^{T}$, which corresponds to eigenvalue 1. Finally, the weights in Equations~\ref{eq:mixture} and~\ref{eq:probs} are replaced by $\pi_k$, $k=1,\ldots,K$. For an overview of hidden Markov model theory and applications the reader is referred to \citet{hmm1,Fruhwirth:06}.

The fetal lamb data \citep{leroux} consists of $n=240$ body movement measurements of a fetal lamb at consecutive 5 second intervals. The dependence of consecutive measurements is a sensible assumption here: a measurement at a specific intensity is quite likely to be followed by a similar one, as displayed in the corresponding time series in Figure \ref{fig:hist} (right). Fr{\"u}hwirth-Schnatter (\citeyear{Fruhwirth:01, Fruhwirth:06}) modelled this time series as a Poisson process where the intensity changes according to a $K$-state hidden Markov process, that is,
\begin{eqnarray*}
x_i|z_i&\sim& \mathcal P(\lambda_{z_i}), \mbox{ independent for}\quad i=1,\ldots,n \\
P(z_i = k|z_{i-1} = \ell)&=& w_{\ell k}, \quad k,\ell\in\{1,\ldots,K\},\quad i=2,\ldots,n,
\end{eqnarray*}
with $\sum_{k=1}^{K}w_{\ell k} = 1$ for all $\ell = 1,\ldots,K$, given some initial distribution $(\pi^{(0)}_1,\ldots,\pi^{(0)}_K)$ for $i = 1$. Given $K$, the parameters to be estimated is the vector of intensities $\boldsymbol{\lambda} = (\lambda_1,\ldots,\lambda_K)$ and the matrix of state-transition probabilities $\boldsymbol w = (w_{\ell k})$, $\ell,k\in\{1,\ldots,K\}$. The number of states is estimated between 3 and 4. We will assume that $K=4$ and run a Gibbs sampler for $m = 10000$ iterations after a burn-in period of $10000$, using the same prior assumptions as discussed in \citet{Fruhwirth:01}.

At first we define the complete likelihood of the model, needed by  \code{sjw} and \code{ecr} methods. The function \code{complete.hmm.poisson.loglikelihood} accepts as input the observed count data (\code{x}), a vector of allocations (\code{z}) and an $K\times J$ dimensional array (\code{pars}) of parameters with $J=K+2$, where: $j=1$ corresponds to Poisson means ($\boldsymbol\lambda$), $j=2,\ldots,K+1$ corresponds to $K$ the columns of the state transition matrix $\boldsymbol w$ and $j=K+2$ corresponds to the left eigenvectors of $\boldsymbol w$. Note here that given $\boldsymbol w$ it is not necessary to save the eigenvectors, but this will make the computation of the complete log-likelihood a little bit faster. The value returned corresponds to the logarithm of Equation~\ref{eq:complete2}.

\begin{Sinput}
R> complete.hmm.poisson.loglikelihood <- function(x, z, pars){
+    post <- sum(dpois(x, pars[z, 1], log = T))
+    logprobs <- log(pars[ , 2:(K+1)])
+    ev <- pars[ , K+2]
+    for(i in 2:n){
+    post <- post + logprobs[z[i-1], z[i]]}
+    post <- post + log(ev[z[1]])
+    return(post)}
\end{Sinput} 

\begin{figure}
\begin{tabular}{ccc}
\includegraphics[width=.3\textwidth]{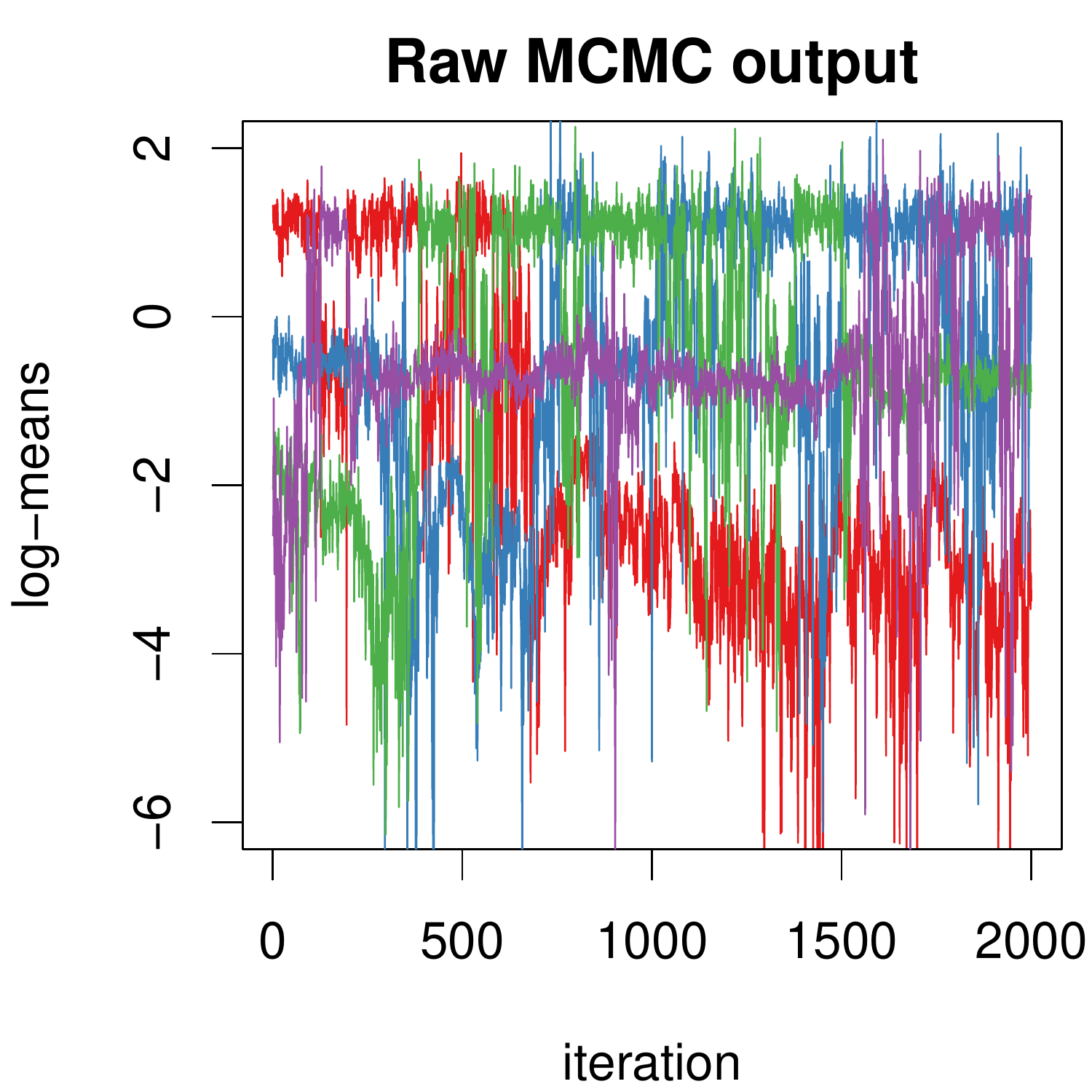}&
\includegraphics[width=.3\textwidth]{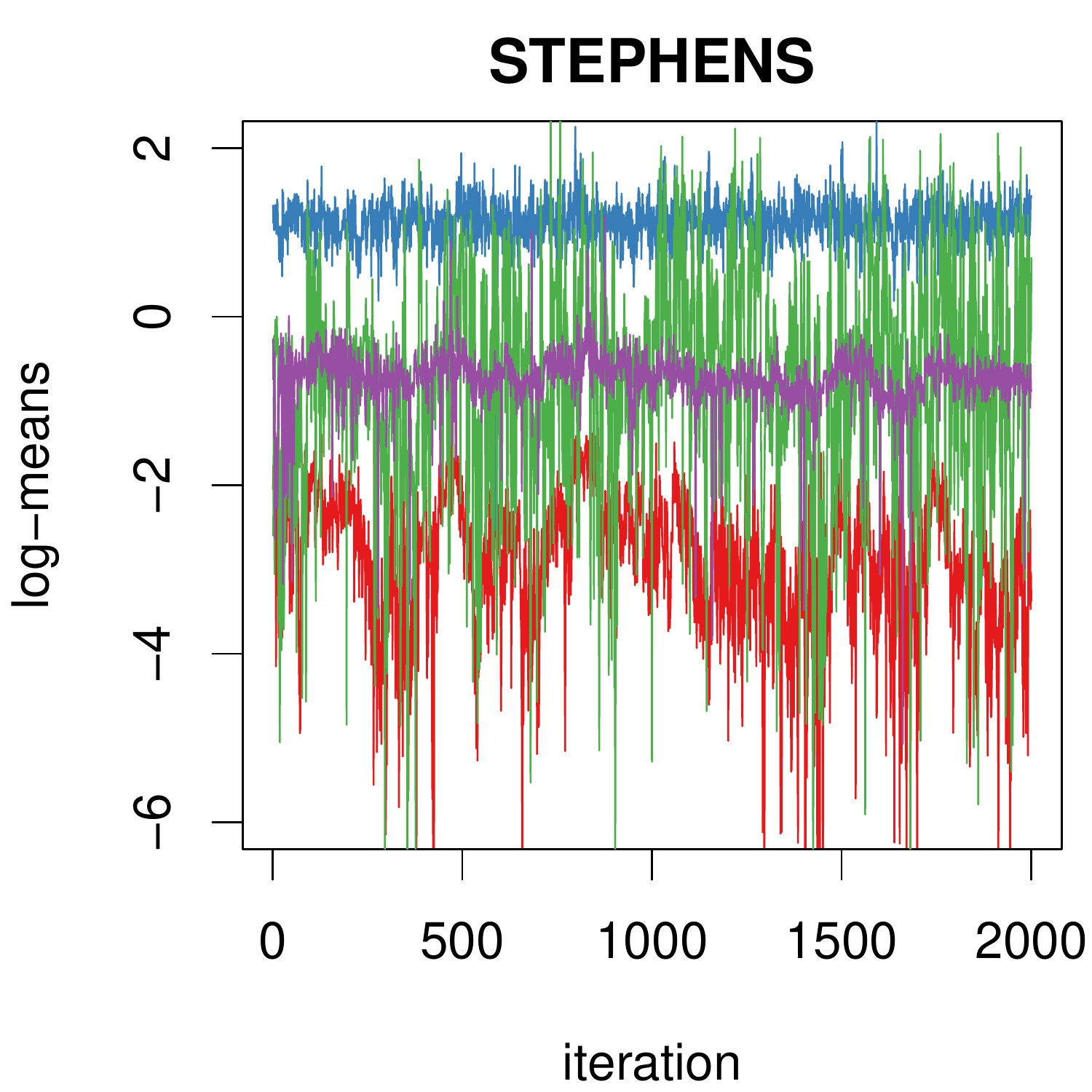}&
\includegraphics[width=.3\textwidth]{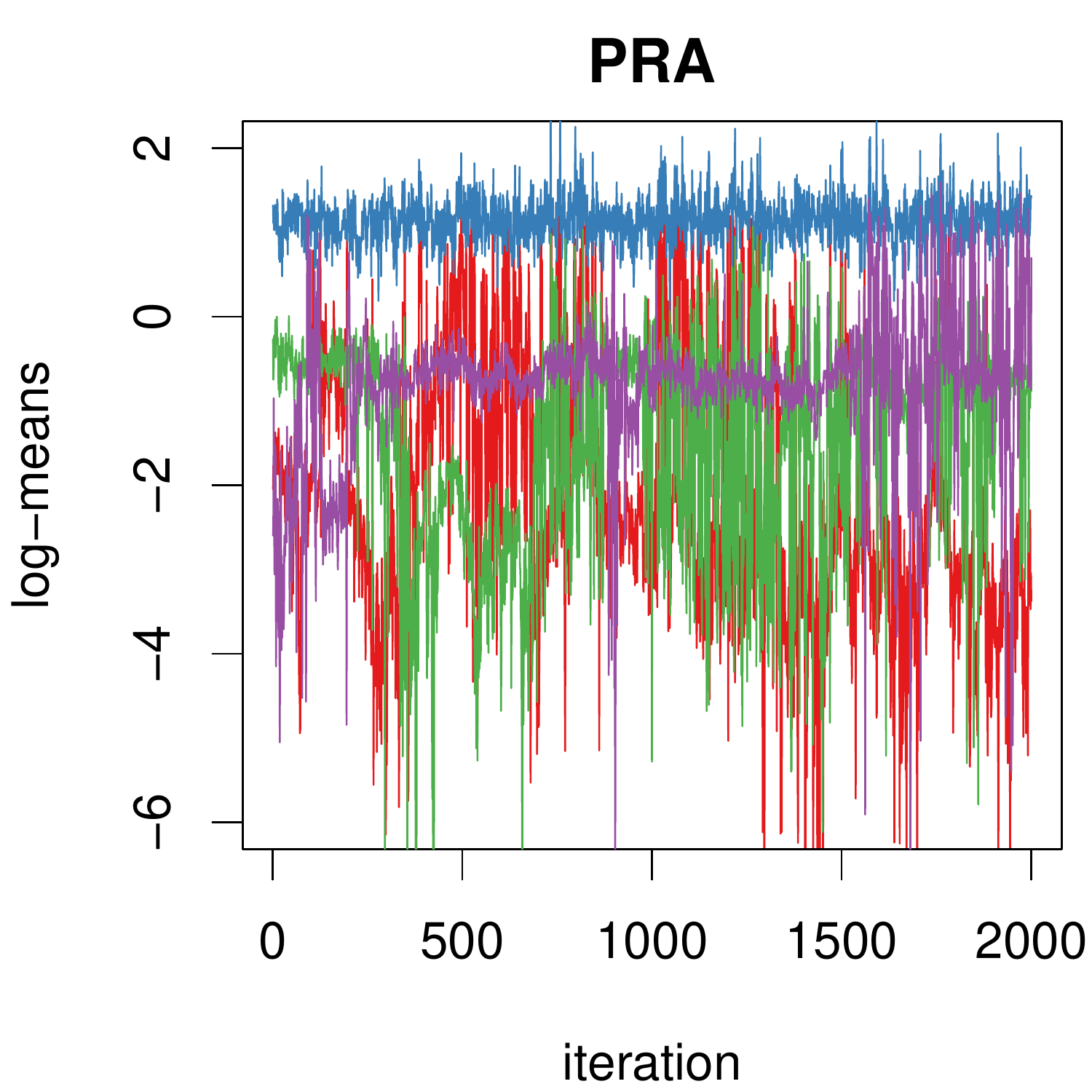}\\
\hspace{6.5ex}(a)&\hspace{6.5ex}(b)&\hspace{6.5ex}(c)\\
\includegraphics[width=.3\textwidth]{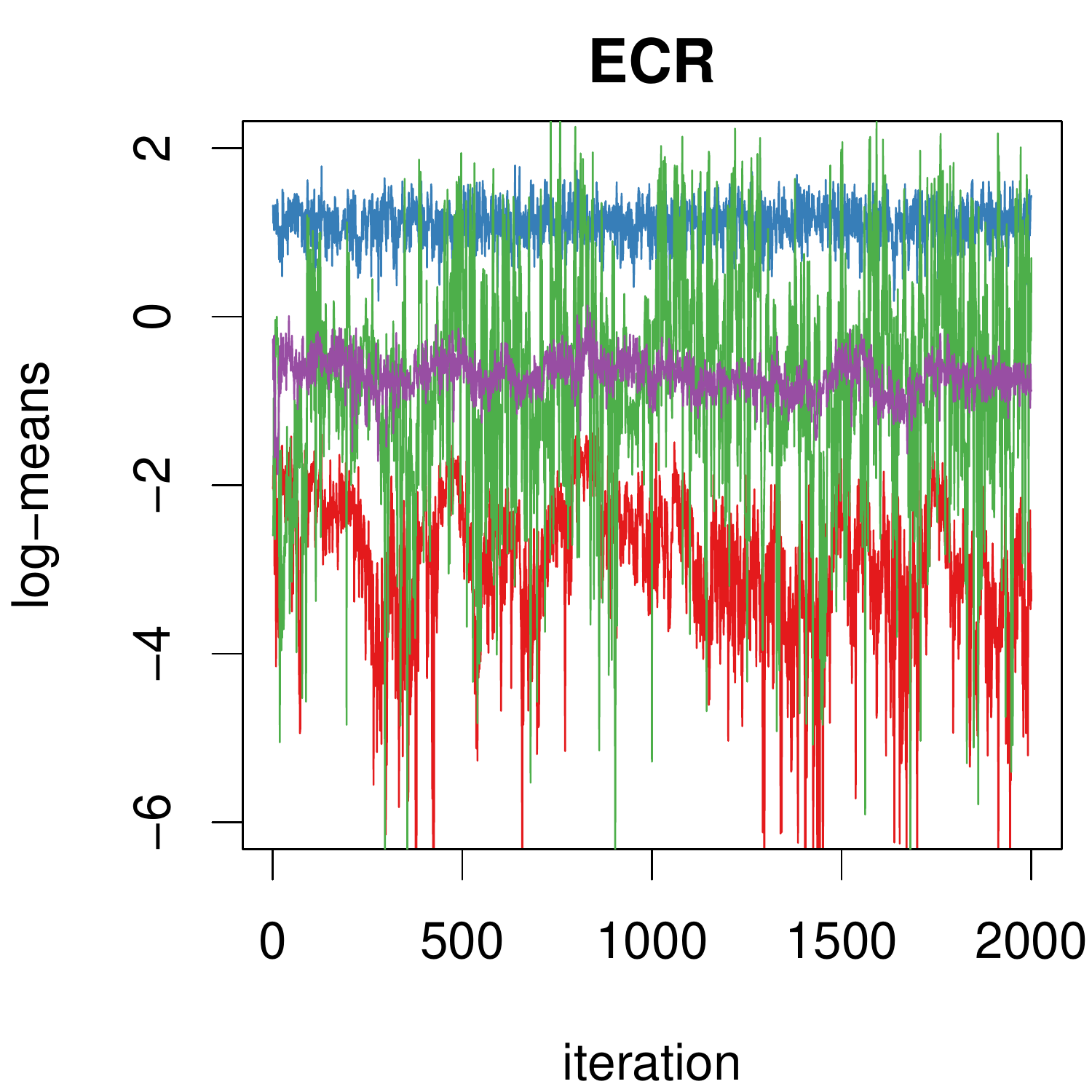}&
\includegraphics[width=.3\textwidth]{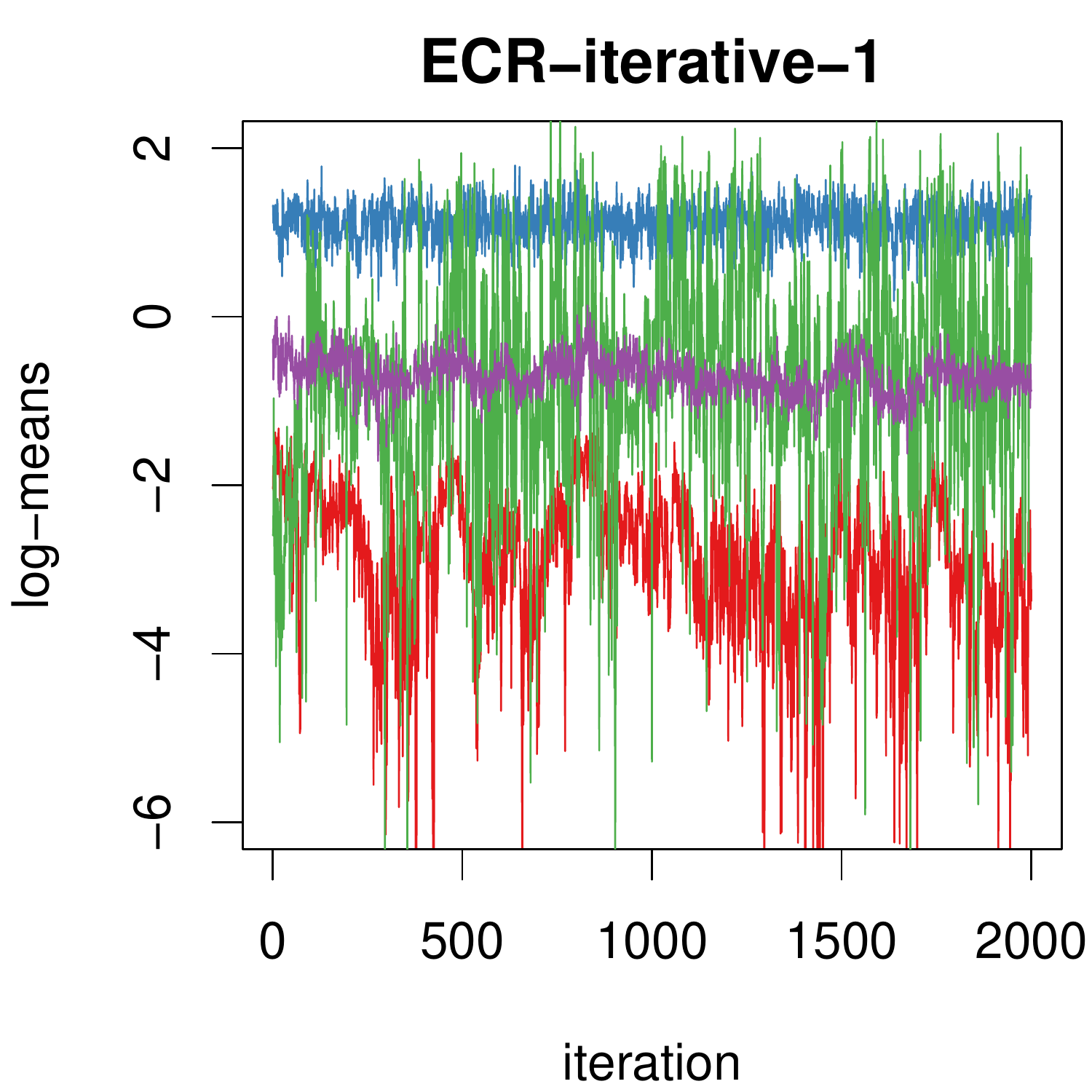}&
\includegraphics[width=.3\textwidth]{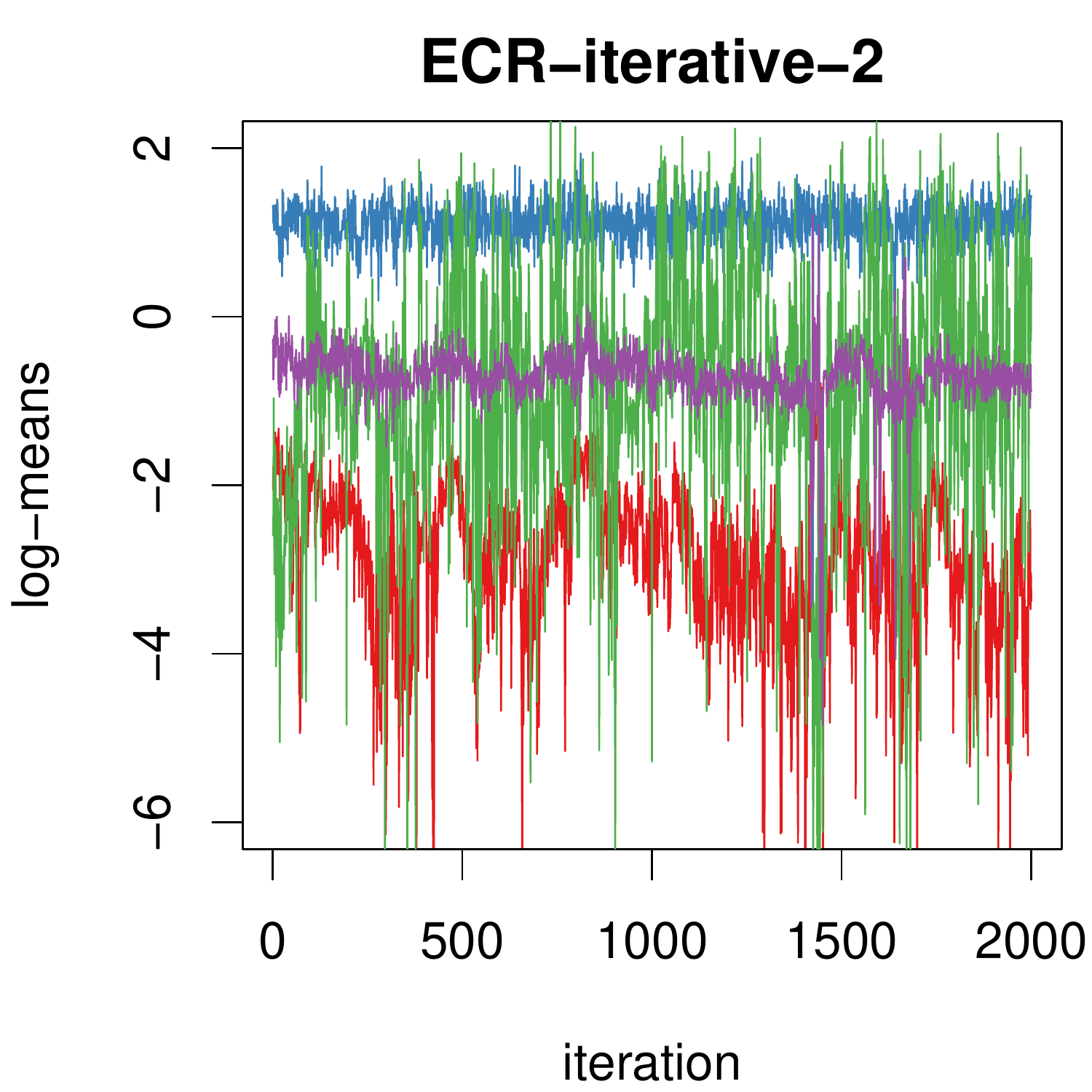}\\
\hspace{6.5ex}(d)&\hspace{6.5ex}(e)&\hspace{6.5ex}(f)\\
\includegraphics[width=.3\textwidth]{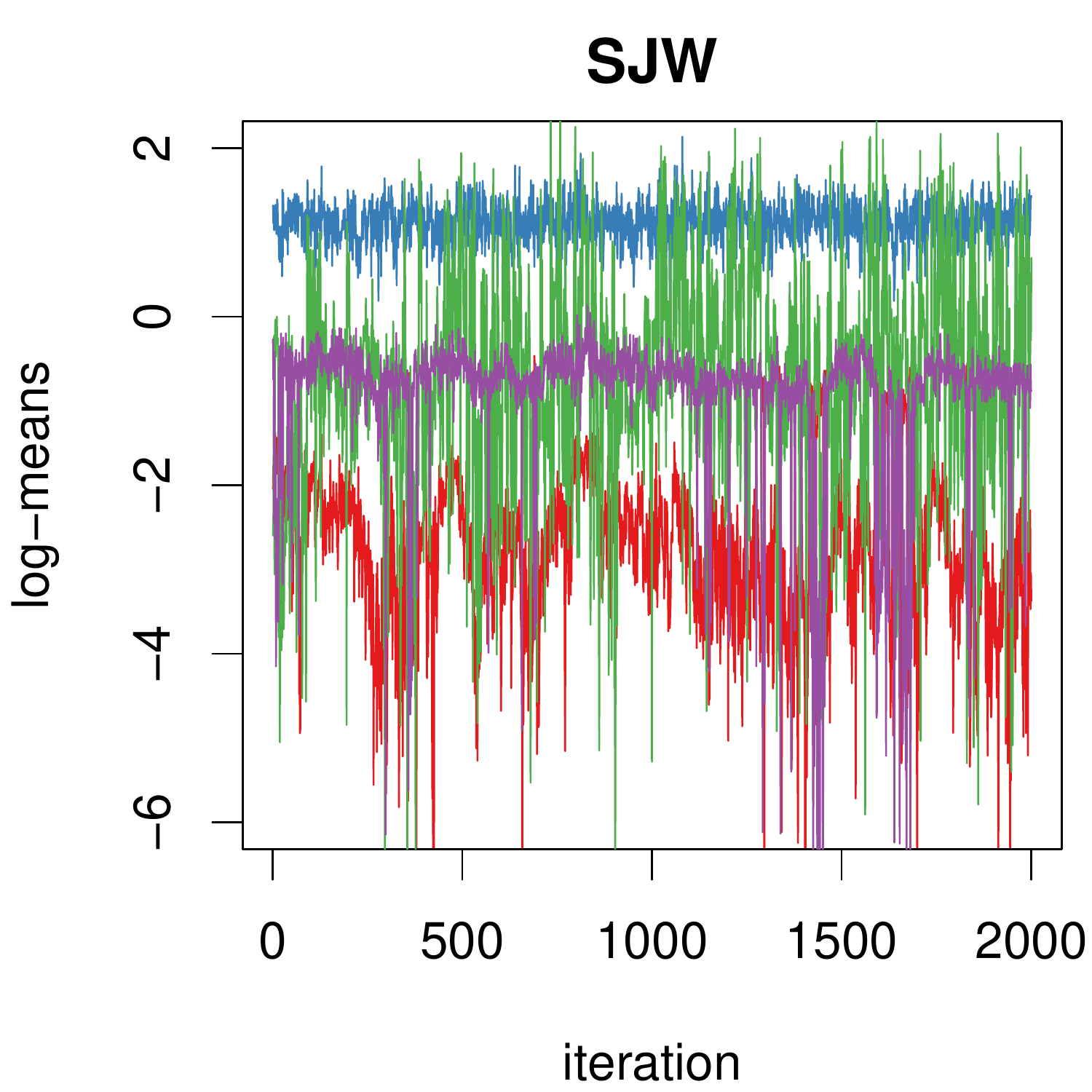}&
\includegraphics[width=.3\textwidth]{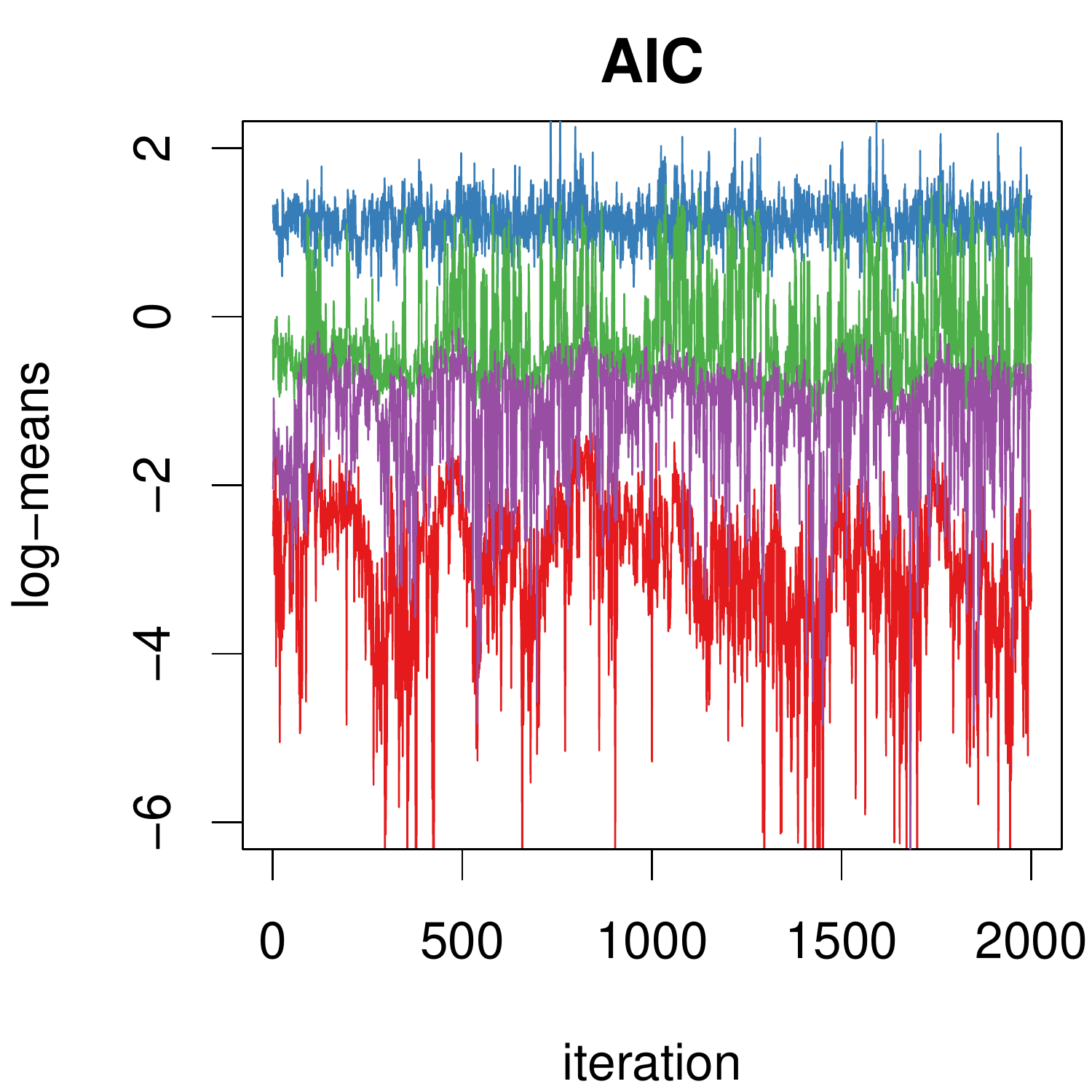}&
\includegraphics[width=.3\textwidth]{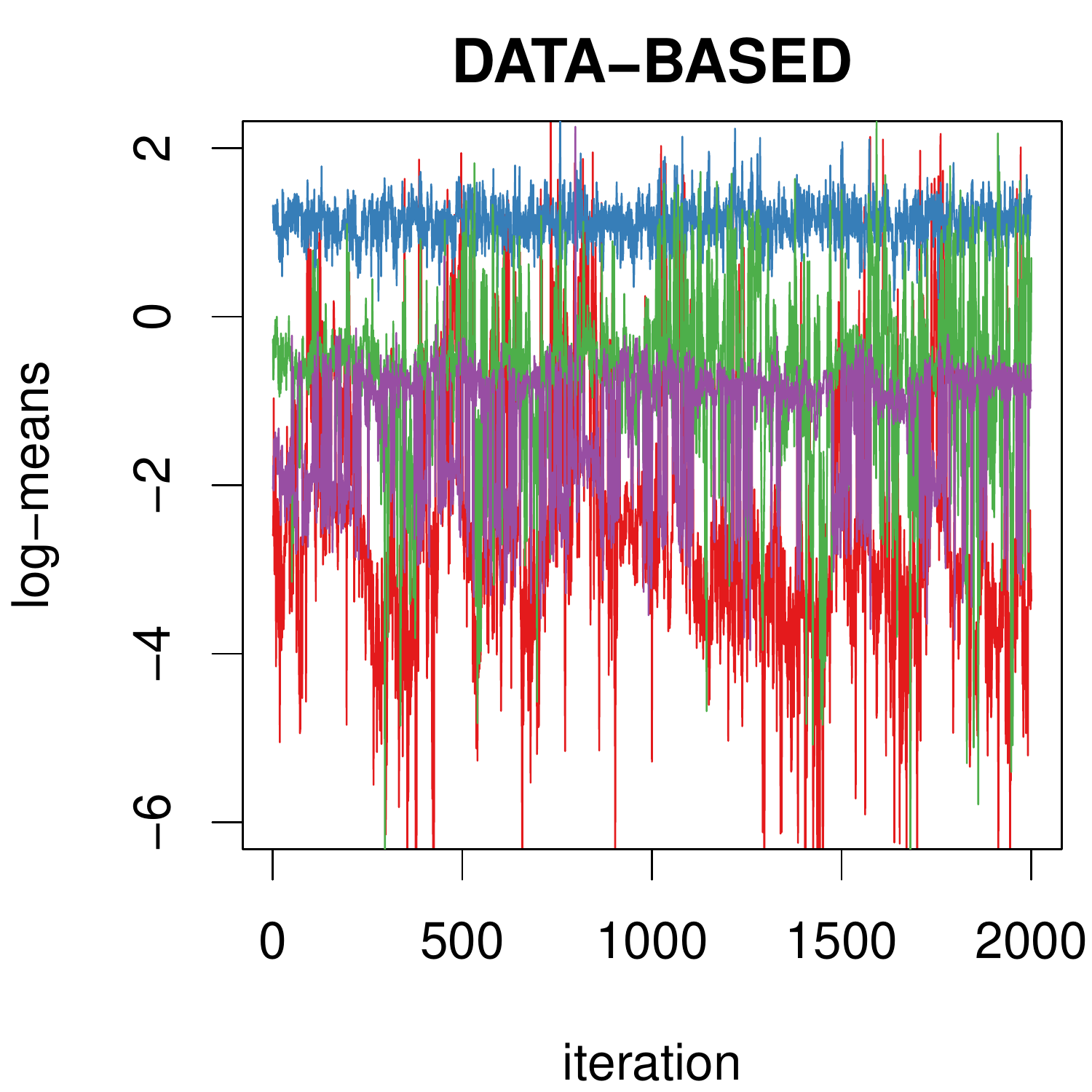}\\
\hspace{6.5ex}(g)&\hspace{6.5ex}(h)&\hspace{6.5ex}(i)
\end{tabular}
\caption{Lamb data ($K=4$). (a): Raw MCMC sample of $\log\lambda_{k}$, $k=1,\ldots,4$. (b), (c), (d), (e), (f), (g), (h), (i): Reordered values by applying the permutations returned by \code{label.switching} function, according to methods: \code{stephens}, \code{pra}, \code{ecr}, \code{ecr-iter-1}, \code{ecr-iter-2}, \code{sjw}, \code{aic} and \code{dataBased} respectively.}\label{fig:lamb}
\end{figure}

The full dataset is available at the \pkg{label.switching} package, while the MCMC sampler is provided as supplemental material. In the end, we have saved the MCMC output to an $m\times K\times J$ array (\code{mcmc.pars}). The raw MCMC output of $\log\lambda_{k}$, $k=1, \ldots,4$ is shown at Figure \ref{fig:lamb}.(a) (every 5th iteration displayed) where the label switching phenomenon is vividly illustrated. Next we apply the relabelling algorithms using the following command.
\begin{Sinput}
R> set <- c("STEPHENS", "PRA", "ECR", "ECR-ITERATIVE-1", "ECR-ITERATIVE-2",
     "SJW", "AIC", "DATA-BASED")
R> ls <- label.switching(method = set, zpivot = z[mapindex, ], z = z, K = 4,
+    prapivot = mcmc.pars[mapindex, , ], p = p, mcmc = mcmc.pars, data = x,
+    complete = complete.hmm.poisson.loglikelihood, constraint = 1)
\end{Sinput}
Note that for the default version of ECR and PRA algorithms we provided the pivot that correspond to iteration \code{mapindex = 3258}, that is, the allocation and parameters that correspond to the iteration that the maximum value of the complete likelihood was observed. After applying the function \code{permute.mcmc} using the resulting permutations, the reordered output of $\log\lambda_k$, $k=1,\ldots,4$ is shown at Figures \ref{fig:lamb}(b)-(i). We conclude that almost all methods suggest that the values of the green-coloured component ($\lambda_3$) have a very large variance compared to the rest. However, this is not the case for \code{pra}, \code{aic}, and to lesser extent for the \code{dataBased} algorithm, where the reordering does not seem to respect the posterior distribution topology.
The proportion of matching allocations between  the available methods is returned by \code{ls\$similarity} and they are shown at the upper diagonal of Table \ref{tab:fish-sim}. Finally, the CPU time per method is shown at second row of Table \ref{tab:fish-times}.

It is interesting to note here that most relabelling algorithms assign no observations (in terms of their single best clusterings) into the third (green) component. In particular, the estimated number of observations assigned to each cluster is:

\begin{Sinput}
R> frequency <- apply(ls$clusters, 1, function(y){freq <- numeric(K);
+    for(j in 1:K){freq[j] = length(which(y == j))}; return(freq)});
+    rownames(frequency) <- 1:K; frequency
\end{Sinput}
\begin{Soutput}
  STEPHENS PRA ECR ECR-ITERATIVE-1 ECR-ITERATIVE-2 SJW AIC DATA-BASED
1      128 117 126             127             132 140 136         90
2        6   6   6               6               6   6   6          6
3        0   6   0               0               0   0  21          0
4      106 111 108             107             102  94  77        144
\end{Soutput}
suggesting that the four-state hidden Markov model might be overparameterized for the fetal lamb dataset.

\subsection{Multivariate normal mixtures}\label{sec:mvn}

In this section, the \pkg{label.switching} package is applied to simulated data from mixtures of multivariate normal distributions. Let $\boldsymbol x_i \in \mathbb R^{d}$ denotes a $d$-dimensional random vector and assume that $$\boldsymbol x_i\sim \sum_{k=1}^{K}w_k\mathcal N_d(\boldsymbol\mu_k,\boldsymbol\Sigma_k),$$
independent for $i=1,\ldots,n$, with $\boldsymbol \mu_k\in \mathbb R^{d}$ and $\boldsymbol \Sigma_k\in\mathcal M^{d\times d}$, $k=1,\ldots,K$, where $\mathcal M^{d\times d}$ denotes the space of $d\times d$ positive definite matrices. Let $\boldsymbol\Lambda_k:=\boldsymbol\Sigma_k^{-1}$ and a priori assume a normal-Wishart prior distribution $$\boldsymbol\mu_k,\boldsymbol\Lambda_k|\boldsymbol\mu_0,\beta,\boldsymbol W,\nu\sim\mathcal N(\boldsymbol\mu_k|\boldsymbol\mu_0,(\beta\boldsymbol\Lambda_k)^{-1})\mathcal W(\boldsymbol\Lambda_k|\boldsymbol W,\nu),$$
independent for $k =1,\ldots,K$, given the constant hyper-parameters $\beta>0$, $\nu>d-1$, $\boldsymbol\mu_0\in\mathbb R^d$ and $\boldsymbol W\in\mathcal M^{d\times d}$. The mixture weights are a priori distributed according to a non-informative Dirichlet distribution. The reader is referred to the supplementary material for the details of the hyper-parameters of the prior distributions.

\begin{table}[t]
\centering
\begin{tabular}{rrrrrrrrrrr}
  \hline
              & \code{steph} & \code{pra} & \code{ecr} & \code{ecr-1} & \code{ecr-2} & \code{sjw} &\code{aic} &\code{data} & \code{user} &true\\ 
  \hline
 \code{steph} &       & 1.000 & 1.000 & 1.000 & 1.000 & 1.000 & 0.880 & 1.000 & 1.000 & 0.940 \\ 
      \code{pra} & -     &       & 1.000 & 1.000 & 1.000 & 1.000 & 0.880 & 1.000 & 1.000 & 0.940 \\ 
      \code{ecr} & 0.993 & -     &       & 1.000 & 1.000 & 1.000 & 0.880 & 1.000 & 1.000 & 0.940 \\ 
\code{ecr-1}& 0.996 & -     & 0.989 &       & 1.000 & 1.000 & 0.880 & 1.000 & 1.000 & 0.940 \\ 
\code{ecr-2}& 1.000 & -     & 0.993 & 0.996 &       & 1.000 & 0.880 & 1.000 & 1.000 & 0.940\\ 
       \code{sjw}& -     & -     & -     & -     & -     &       & 0.880 & 1.000 & 1.000 & 0.940 \\ 
       \code{aic}& 0.896 & -     & 0.889 & 0.889 & 0.896 & -     &       & 1.000 & 0.880 & 0.830\\
   \code{data}& 1.000 & -     & 0.993 & 0.996 & 1.000 & -     & 0.896 &       & 1.000 & 0.940\\
    \code{user}  & 0.971 & -     & 0.971 & 0.968 & 0.971 & -     & 0.892 & 0.971 &       & 0.940 \\
   true          & 0.929 & -     & 0.929 & 0.929 & 0.929 & -     & 0.836 & 0.929 & 0.918 &       \\
   \hline
\end{tabular}
\caption{ \code{ls\$similarity}: Proportion of matching allocations for the single best-clusterings between the relabelling algorithms, the user-defined permutations and the true allocations for the first and second multivariate dataset (upper and lower diagonal, respectively).}\label{tab:mvn}
\end{table}

We simulated two datasets of $n = 100$ and $280$ observations from a bivariate ($d = 2$) mixture  with $K=4$ and $9$ components, respectively. The real values used to generate the first dataset were chosen as $\boldsymbol\mu_k = 2.5\left(\cos\frac{(k-1)\pi}{4},\sin\frac{(k-1)\pi}{4}\right)^{t}$, $w_k = 1/K$, $\Sigma_{11k} = \Sigma_{22k} = 1$, $\Sigma_{12k} = \Sigma_{21k} = 0$, $k = 1,\ldots,4$. The real values for the second dataset were chosen as $\boldsymbol\mu_k = 6\left(\cos\frac{(k-1)\pi}{8},\sin\frac{(k-1)\pi}{8}\right)^{t}$, $\Sigma_{11k} = \Sigma_{22k} = 1$, $\Sigma_{12k} = \Sigma_{21k} = 0$, $w_k = 0.1$, for $k = 1,\ldots,8$ and for the last component: $\mu_{9} = (0,0)$, $\Sigma_{119} = \Sigma_{229} = 4$, $\Sigma_{129} = \Sigma_{219} = 0$ and $w_9 = 0.2$. 

The Gibbs sampler was implemented next, using the package \pkg{mvtnorm} \citep{mvtnorm} in order to simulate from the full conditional distributions.   The source code is provided in the supplementary material (\code{gibbsSampler} function). This functions returns the following objects: \code{mcmc}, \code{MLindex}, \code{z} and \code{p}, which correspond to the simulated parameters, the index which corresponds to the iteration where the maximum value of the complete likelihood is observed, the simulated allocations and the classification probabilities, respectively. More specifically, \code{mcmc}  is an $m\times K\times J$ array, where $J = d + d(d+1)/2 + 1$ denotes the number of different parameter types for the bivariate normal mixture. Hence,  $\mbox{\code{mcmc[t, k, 1]}} = \mu_{1k}^{(t)}$, $\mbox{\code{mcmc[t, k, 2]}} = \mu_{2k}^{(t)}$, $\mbox{\code{mcmc[t, k, 3]}} = \Sigma_{11}^{(t)}$, $\mbox{\code{mcmc[t, k, 4]}} = \Sigma_{22}^{(t)}$, $\mbox{\code{mcmc[t, k, 5]}} = \Sigma_{12}^{(t)}$ and $\mbox{\code{mcmc[t, k, 6]}} = w_k^{(t)}$, $k = 1,\ldots,K$, $t=1,\ldots,m$. The function is called as follows.
\begin{Sinput}
R> gs <- gibbsSampler(iterations = iterations, K = K, x = x, burn = burn)
R> zChain <- gs$z
R> mcmc.pars <- gs$mcmc
R> pivot <- gs$MLindex
R> allocProbs <- gs$p
\end{Sinput} 
For the first dataset we set \code{K = 4}, \code{iterations = 11000} and \code{burn = 1000} and for the second dataset we set \code{K = 9}, \code{iterations = 20000} and \code{burn = 5000}.

All relabelling algorithms are applied to the first dataset, but \code{pra} and \code{sjw} are excluded to the second one due to the large number of permutations $(K!)$ that should be computed for each MCMC iteration. For the \code{sjw} algorithm, the \code{complete.bivariate.normal.loglikelihood}  function (available in the supplementary material) returns the complete log-likelihood function of the bivariate normal mixture. The ordering constraint will be applied to $\mu_{1k}$, $k=1,\ldots,K$. We will also provide an additional set of permutations using the \code{userPerm} option. Assume that after visual inspection of the MCMC draws, the user wishes to check whether the MCMC sample is identifiable by imposing an ordering constraint to $\mu_{1k}-2\mu_{2k}$, $k=1,\ldots,K$. This is easily done using the following commands.
\begin{Sinput}
R> newMCMC <- array(data = NA, dim = c(iterations, K, 7))
R> newMCMC[ , , 1:6] <- mcmc.pars
R> for(k in 1:K){
+ 	newMCMC[ , k, 7] <- mcmc.pars[ , k, 1] - 2*mcmc.pars[ , k, 2]}
R> newConstraint <- aic(newMCMC, constraint = 7) 
\end{Sinput}
Now apply all relabelling algorithms using the \code{label.switching} command, by parsing also the user-defined permutations in order to compare them to the rest of the methods as follows.

\begin{figure}[t]
\begin{tabular}{ccc}
\includegraphics[width=.3\textwidth]{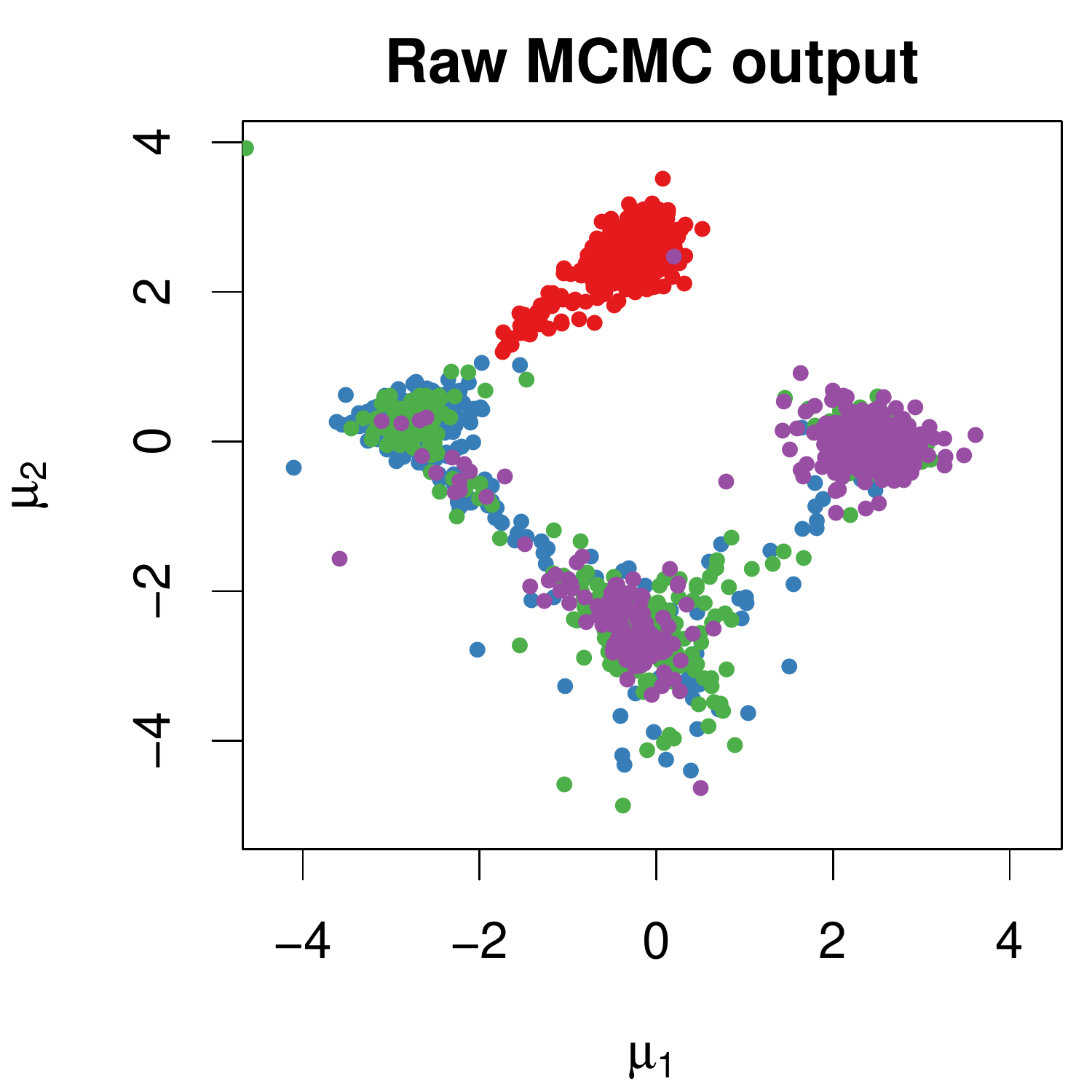}&
\includegraphics[width=.3\textwidth]{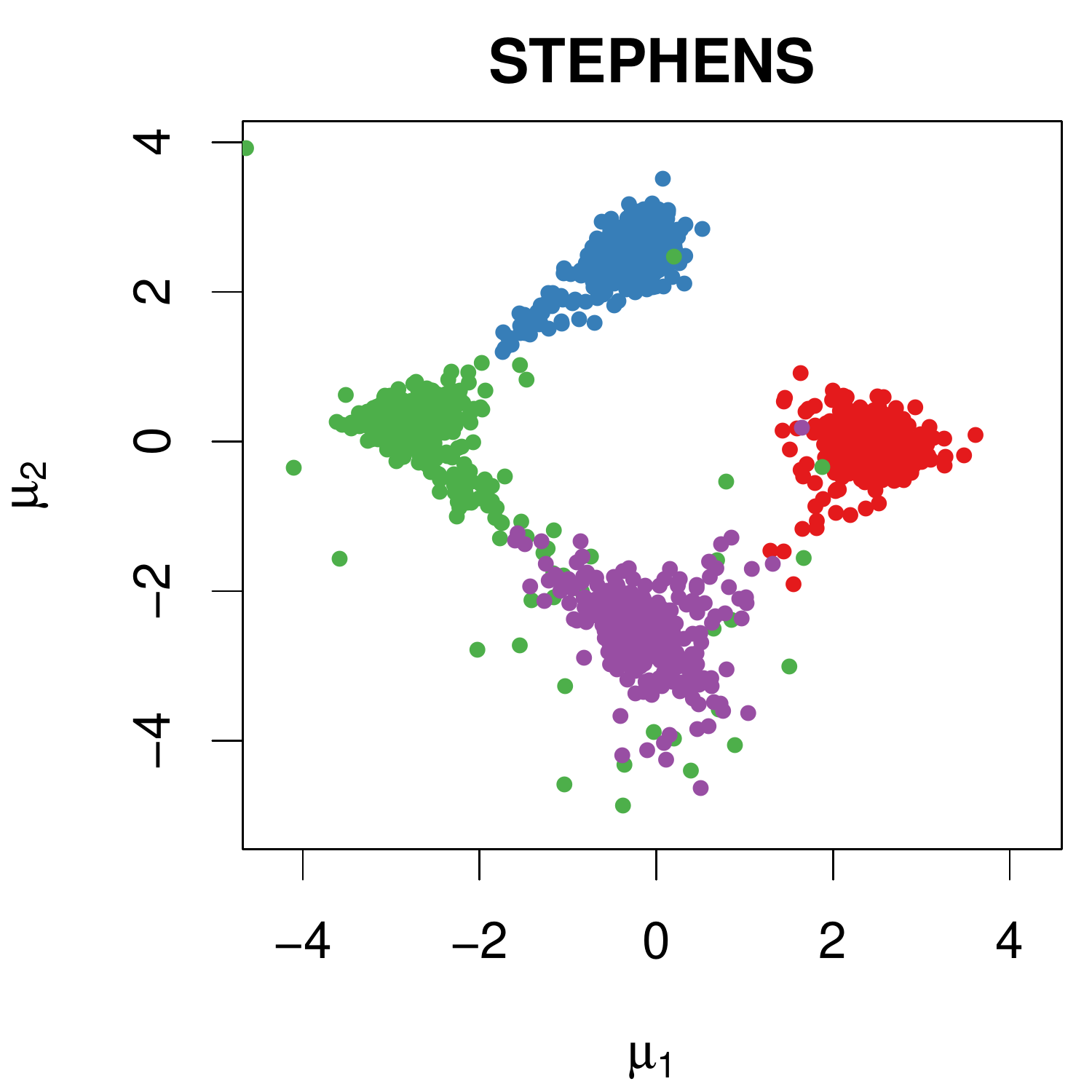}&
\includegraphics[width=.3\textwidth]{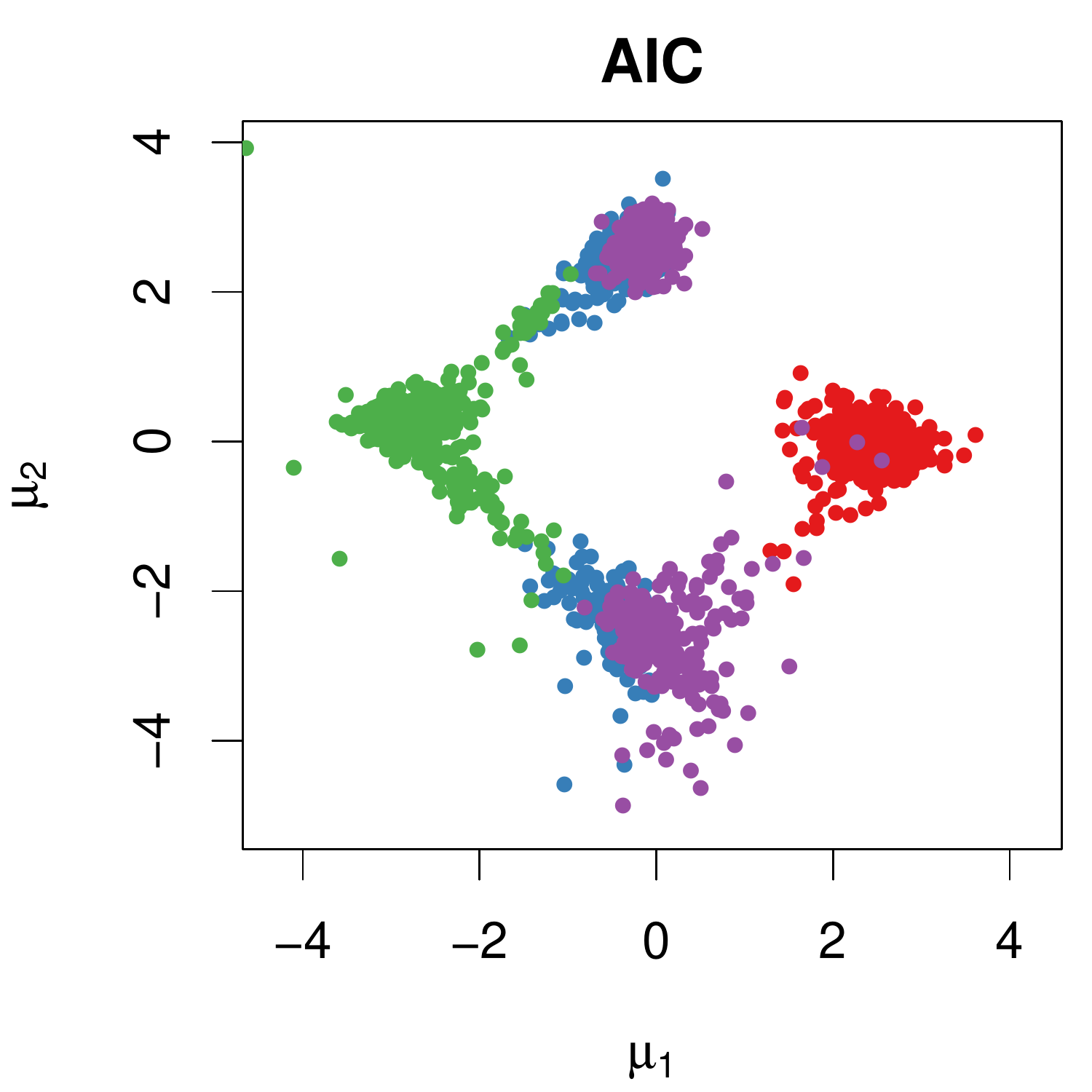}\\
\includegraphics[width=.3\textwidth]{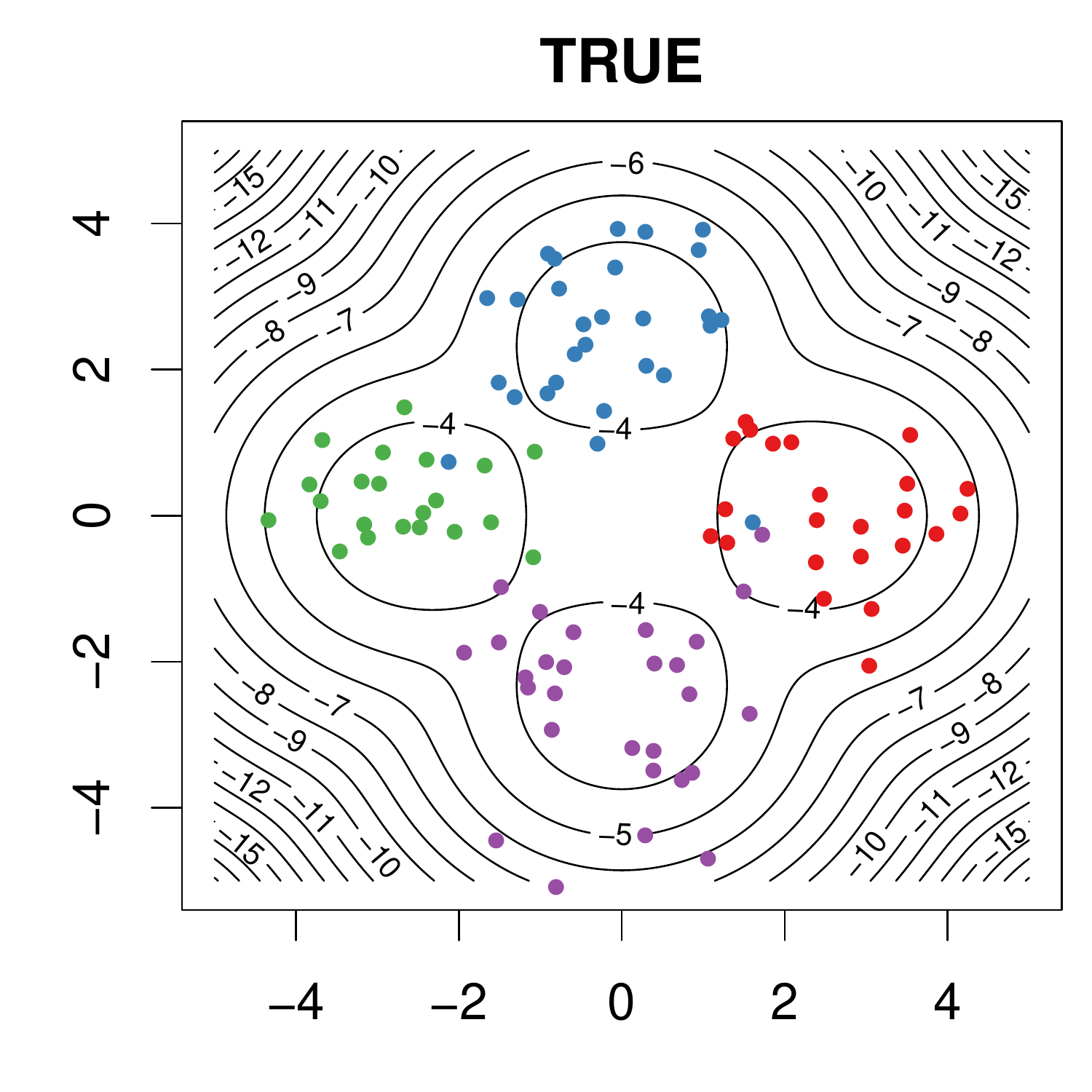}&
\includegraphics[width=.3\textwidth]{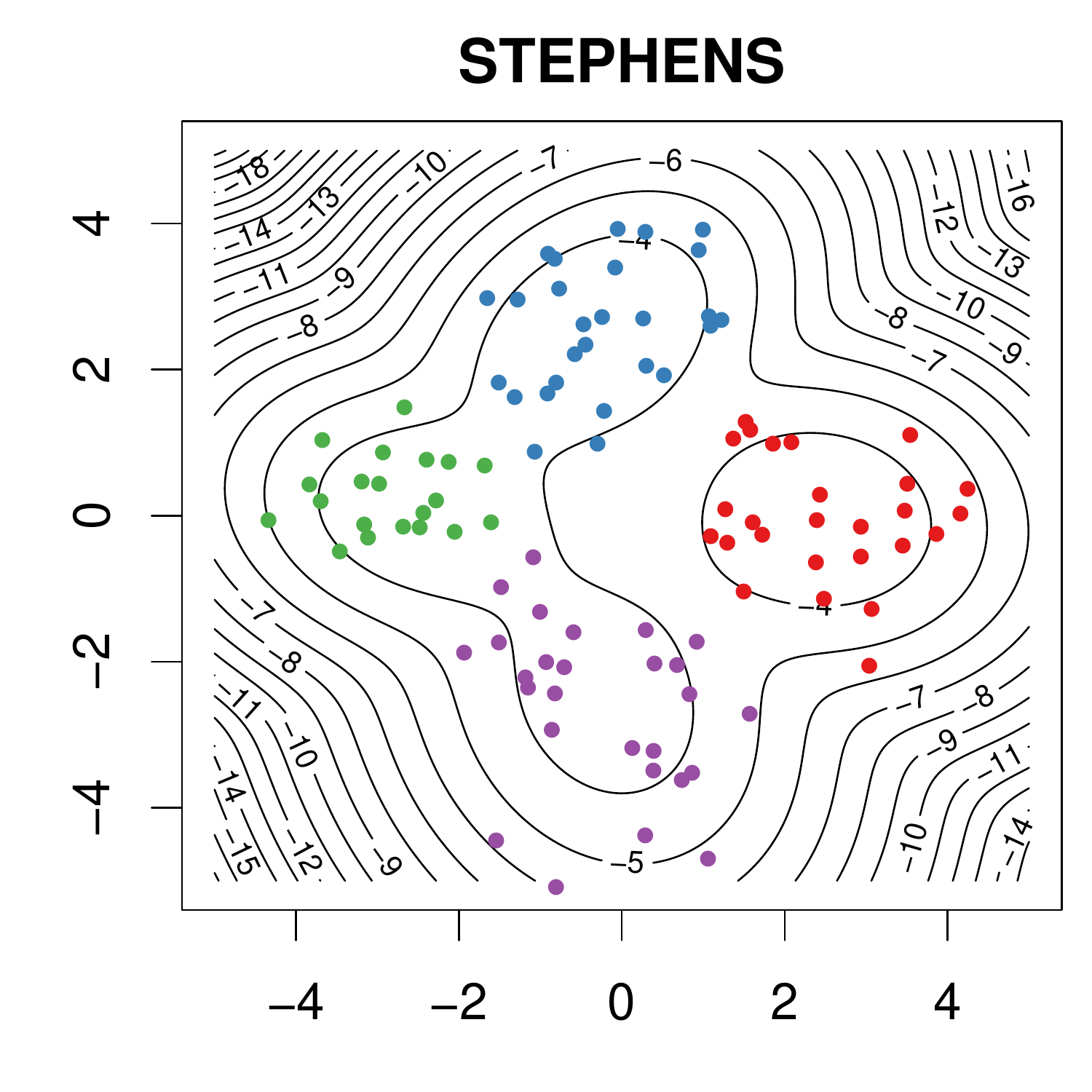}&
\includegraphics[width=.3\textwidth]{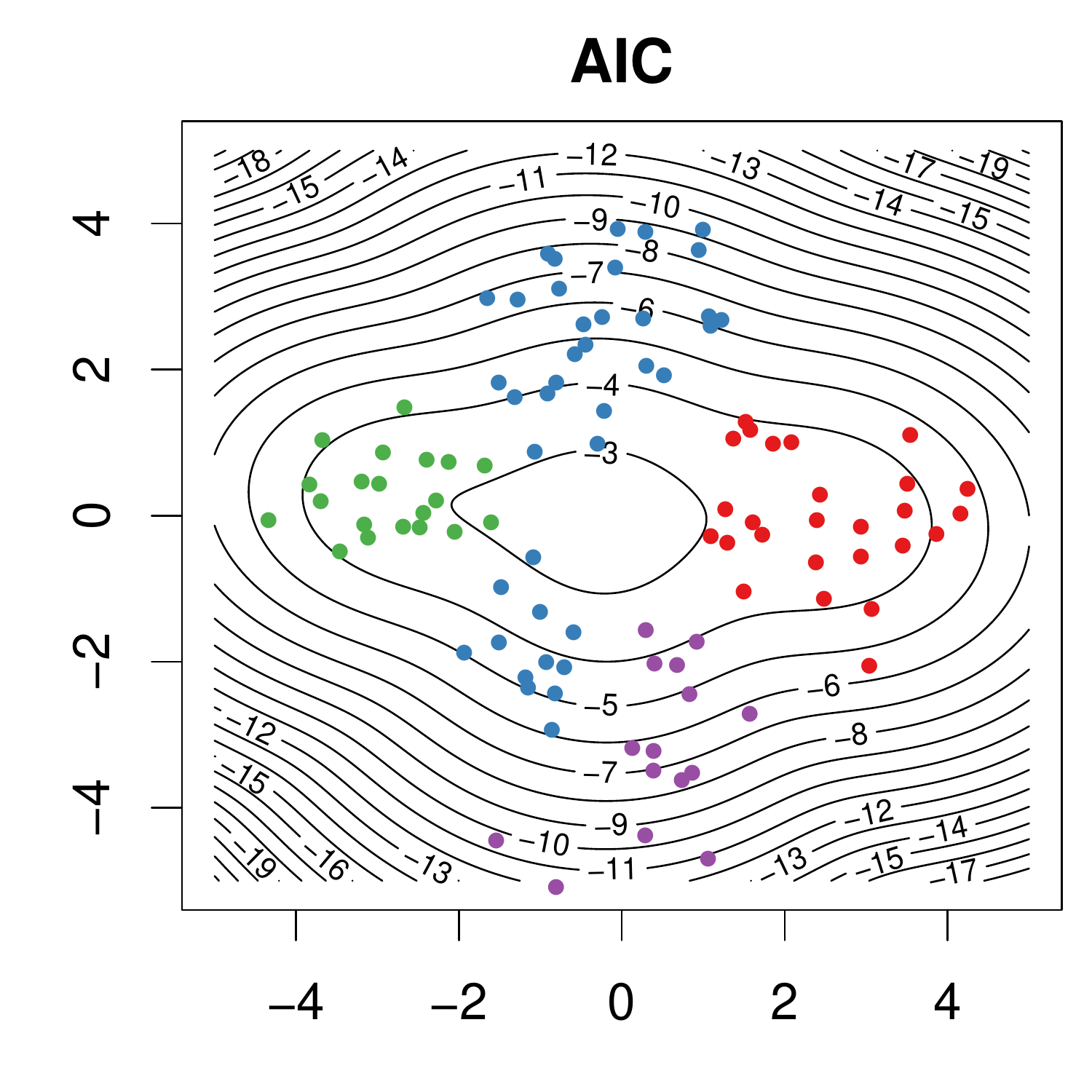}\\
\end{tabular}
\caption{Multivariate dataset 1 ($K=4$). Top: raw MCMC output of $(\mu_1,\mu_2)$ and reordered values according to \code{stephens} and \code{aic} algorithms for a randomly sampled subset of $500$ MCMC iterations. Bottom: True density and clusters of data and corresponding estimates to \code{stephens} and \code{aic} algorithms.}\label{fig:mvn}
\end{figure}

For the first dataset
\begin{Sinput}
R> set <- c("STEPHENS", "PRA", "ECR", "ECR-ITERATIVE-1", "ECR-ITERATIVE-2",
+    "SJW", "AIC", "DATA-BASED", "USER-PERM")
R> ls <- label.switching(method = set, zpivot = zChain[pivot, ], z = zChain,
+    K = 4, prapivot = mcmc.pars[pivot, , ], p = allocProbs,
+    complete = complete.bivariate.normal.loglikelihood,
+    mcmc = mcmc.pars, data = x, sjwinit = pivot, groundTruth = z.real,
+    userPerm = newConstraint$permutations)
\end{Sinput}
For the second dataset
\begin{Sinput}
R> set <- c("STEPHENS", "ECR", "ECR-ITERATIVE-1", "ECR-ITERATIVE-2",
+    "AIC", "DATA-BASED", "USER-PERM")
R> ls <- label.switching(method = set, zpivot = zChain[pivot, ], z = zChain,
+    K = 9, p = allocProbs, mcmc = mcmc.pars, data = x, groundTruth = z.real,
+    userPerm = newConstraint$permutations)
\end{Sinput}
The run-times for the reordering part of each algorithm is displayed in Table \ref{tab:fish-times} (last two rows). The simulation study allows to compare all relabelling methods against the ground truth used to generate the data. This is simply done by parsing the option \code{groundTruth = z.real} to the \code{label.switching} command (\code{z.real} corresponds to the true allocations of the observations). Hence, all permutations now are rearranged in order to maximize their similarity with \code{z.real}. Table \ref{tab:mvn} displayes the coherence between the single best clusterings as returned by \code{ls\$similarity}. Excluding \code{aic}, there is a strong agreement between the relabelling algorithms and the ground truth, as well as within the relabelling algorithms (note the absolute agreement for the first dataset).

\begin{figure}[t]
\begin{tabular}{ccc}
\includegraphics[width=.3\textwidth]{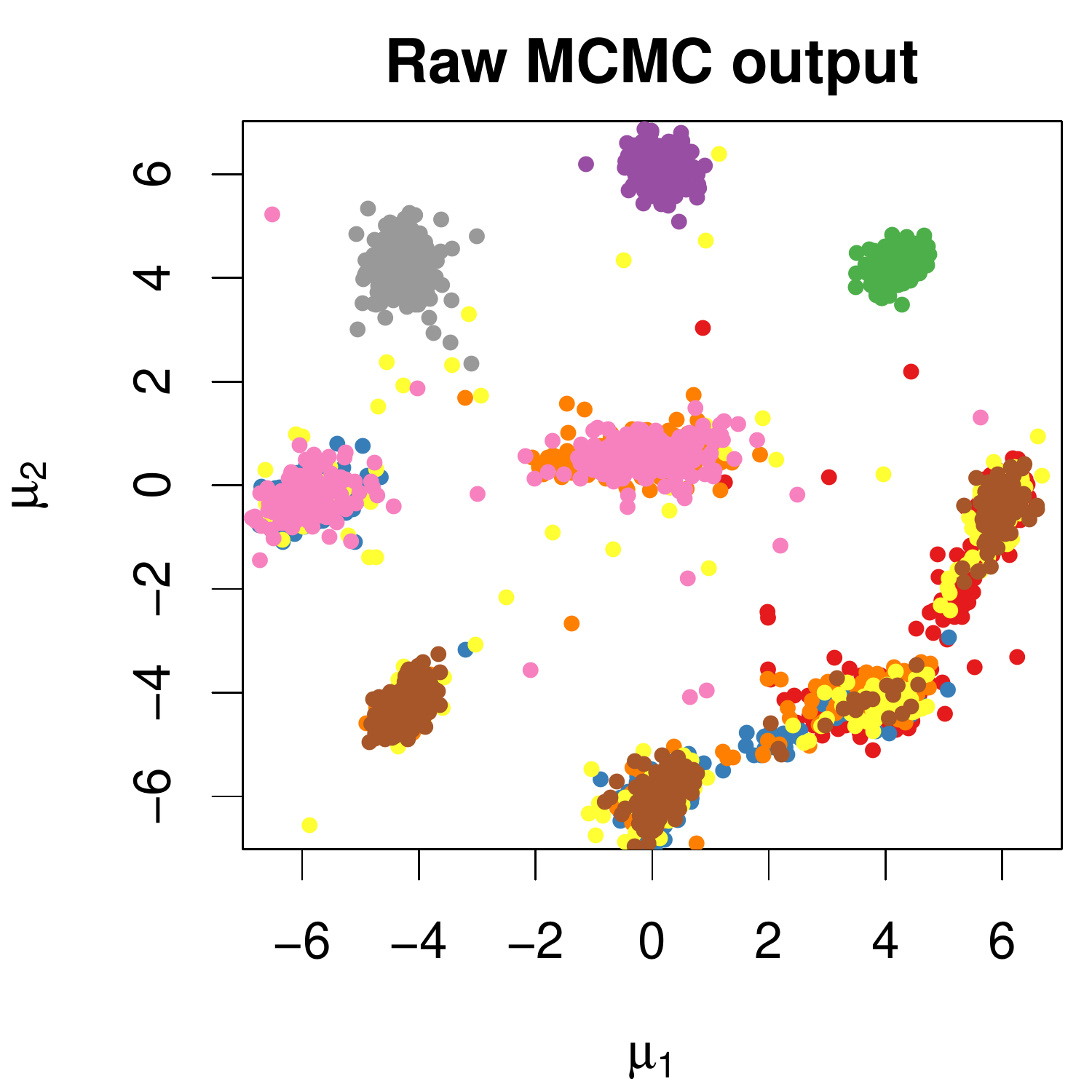}&
\includegraphics[width=.3\textwidth]{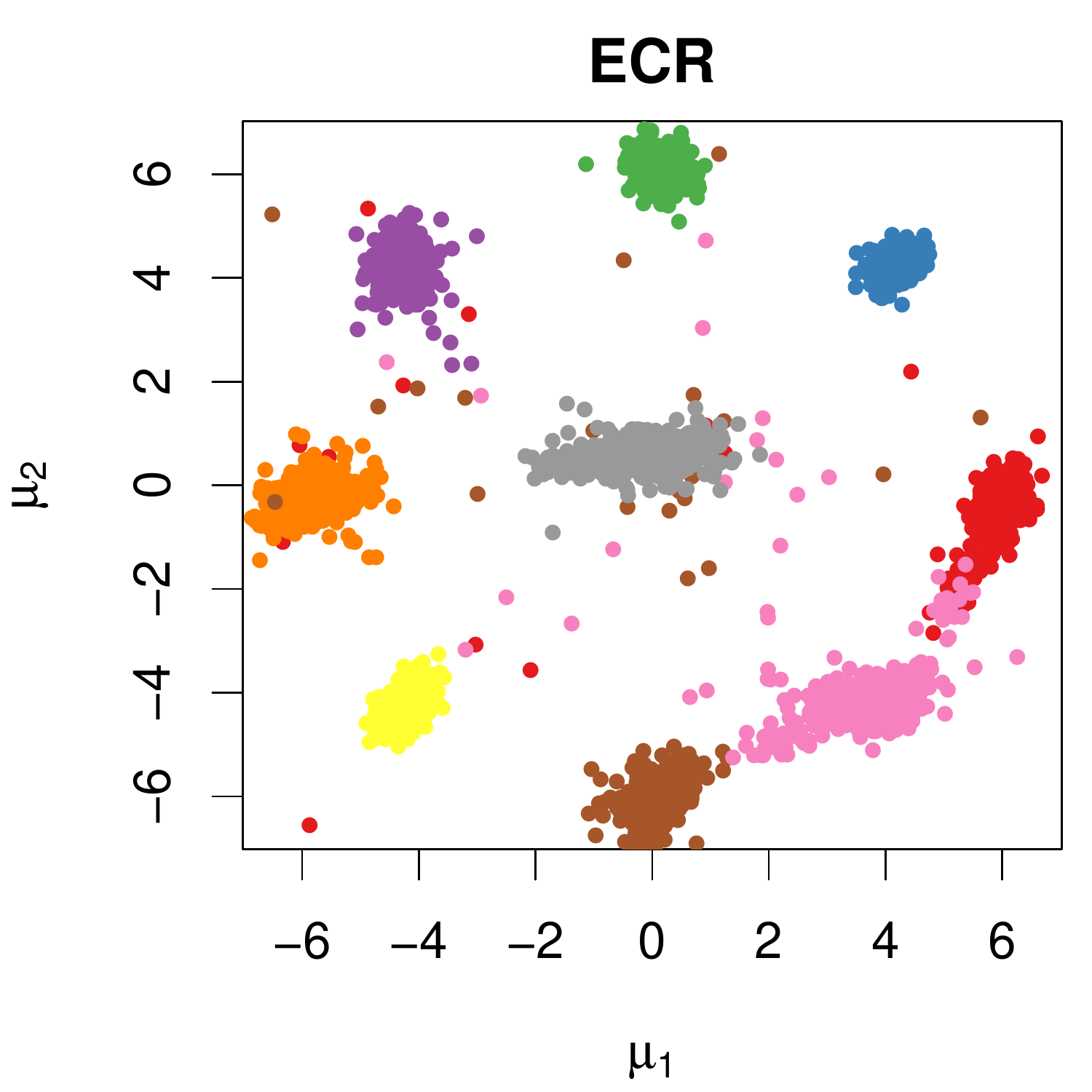}&
\includegraphics[width=.3\textwidth]{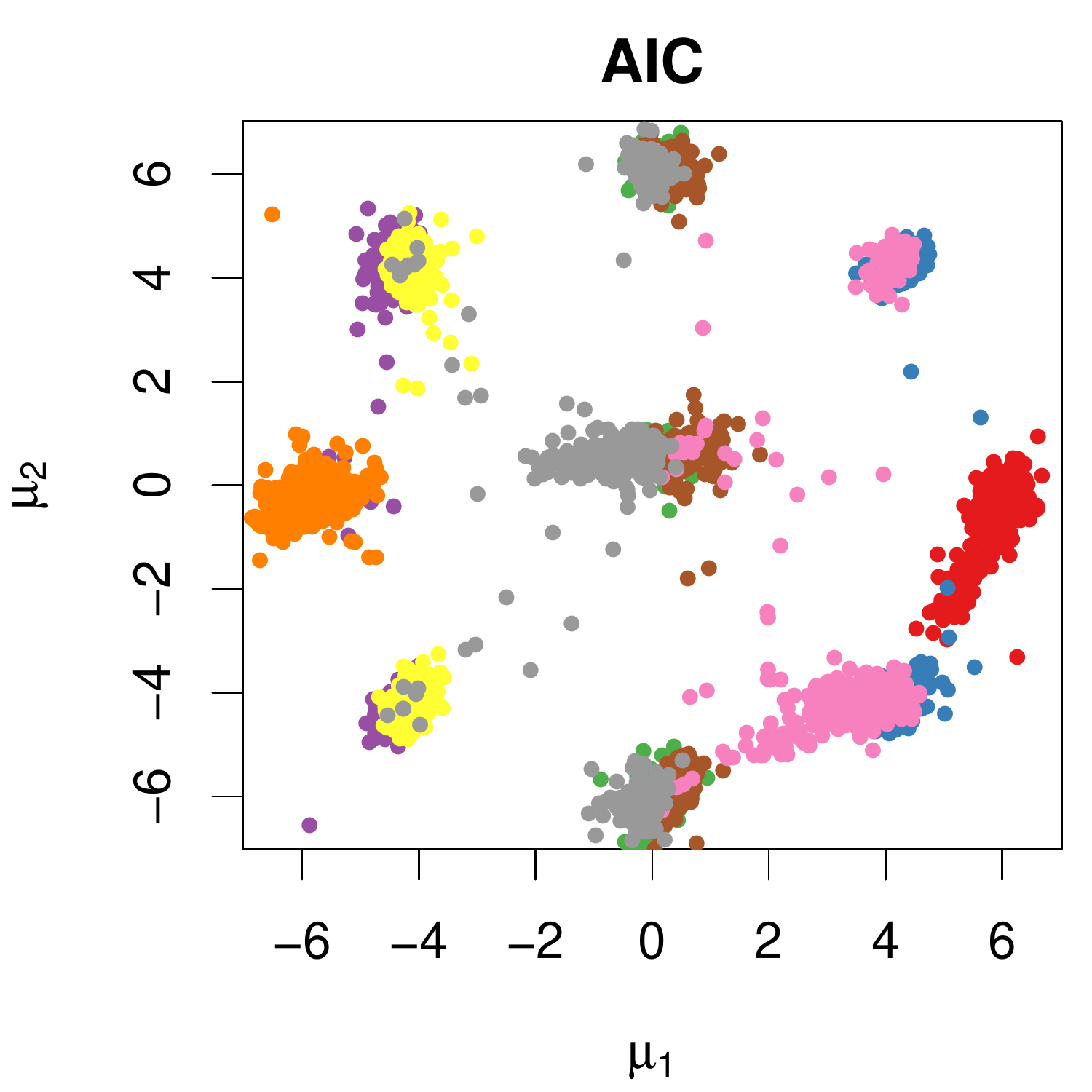}\\
\includegraphics[width=.3\textwidth]{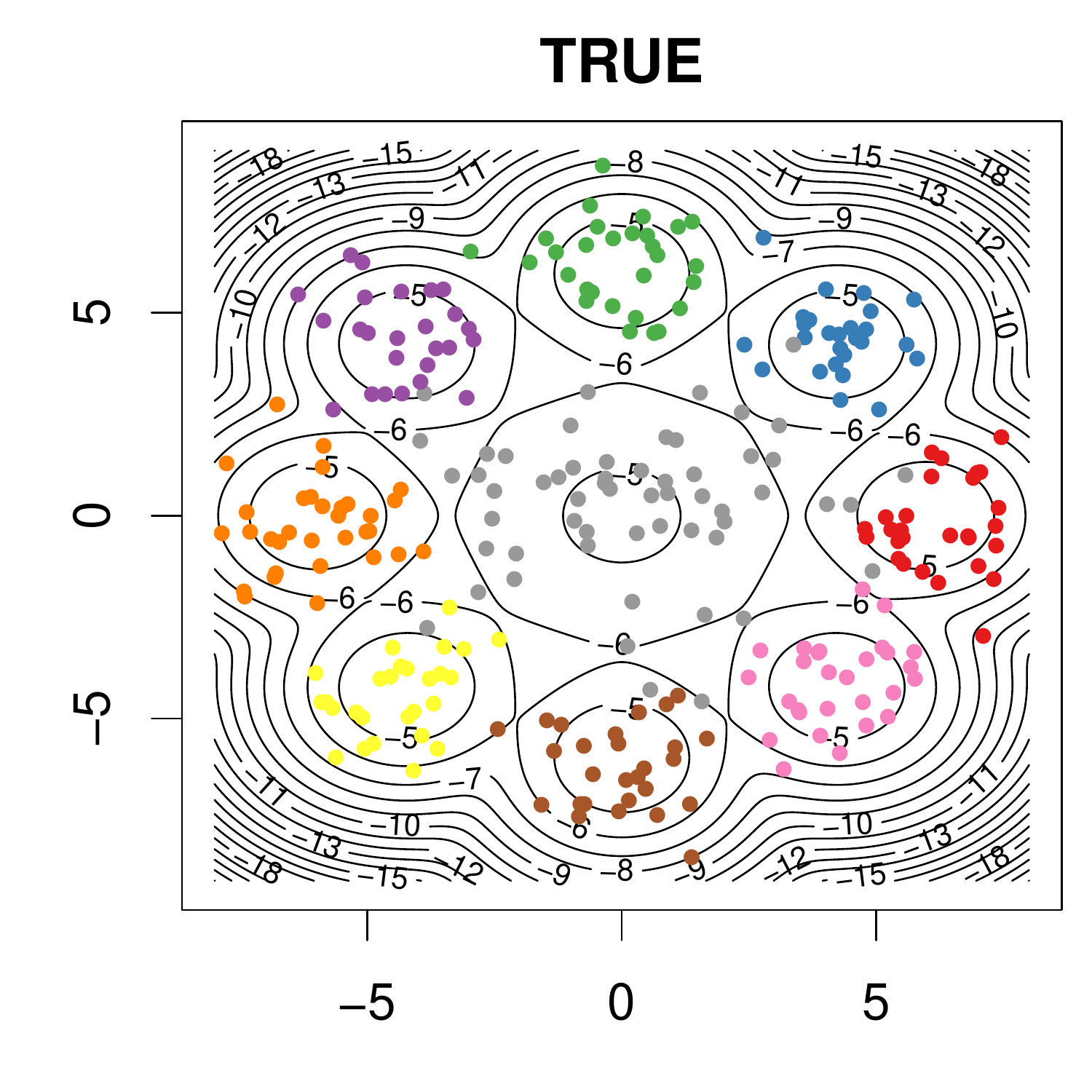}&
\includegraphics[width=.3\textwidth]{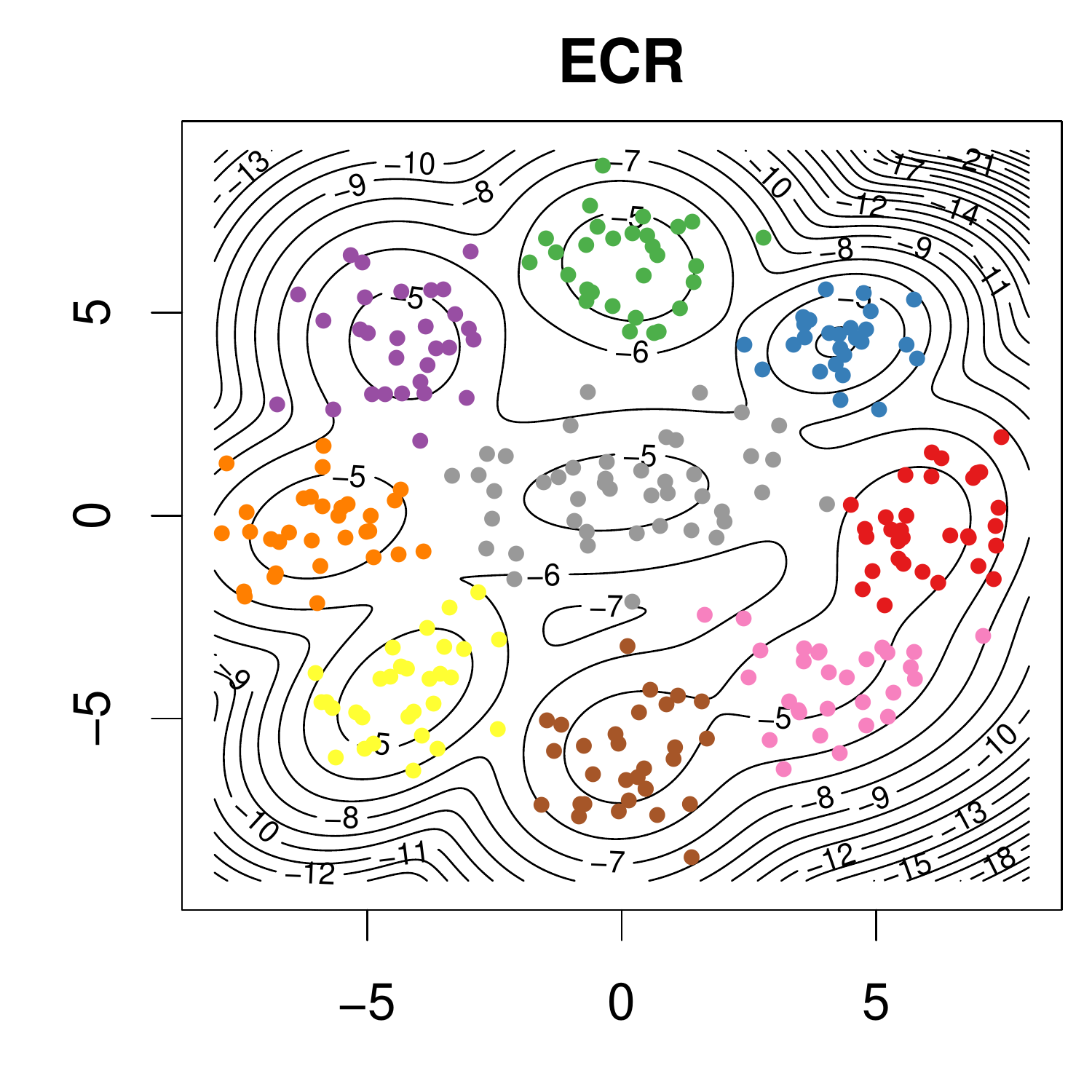}&
\includegraphics[width=.3\textwidth]{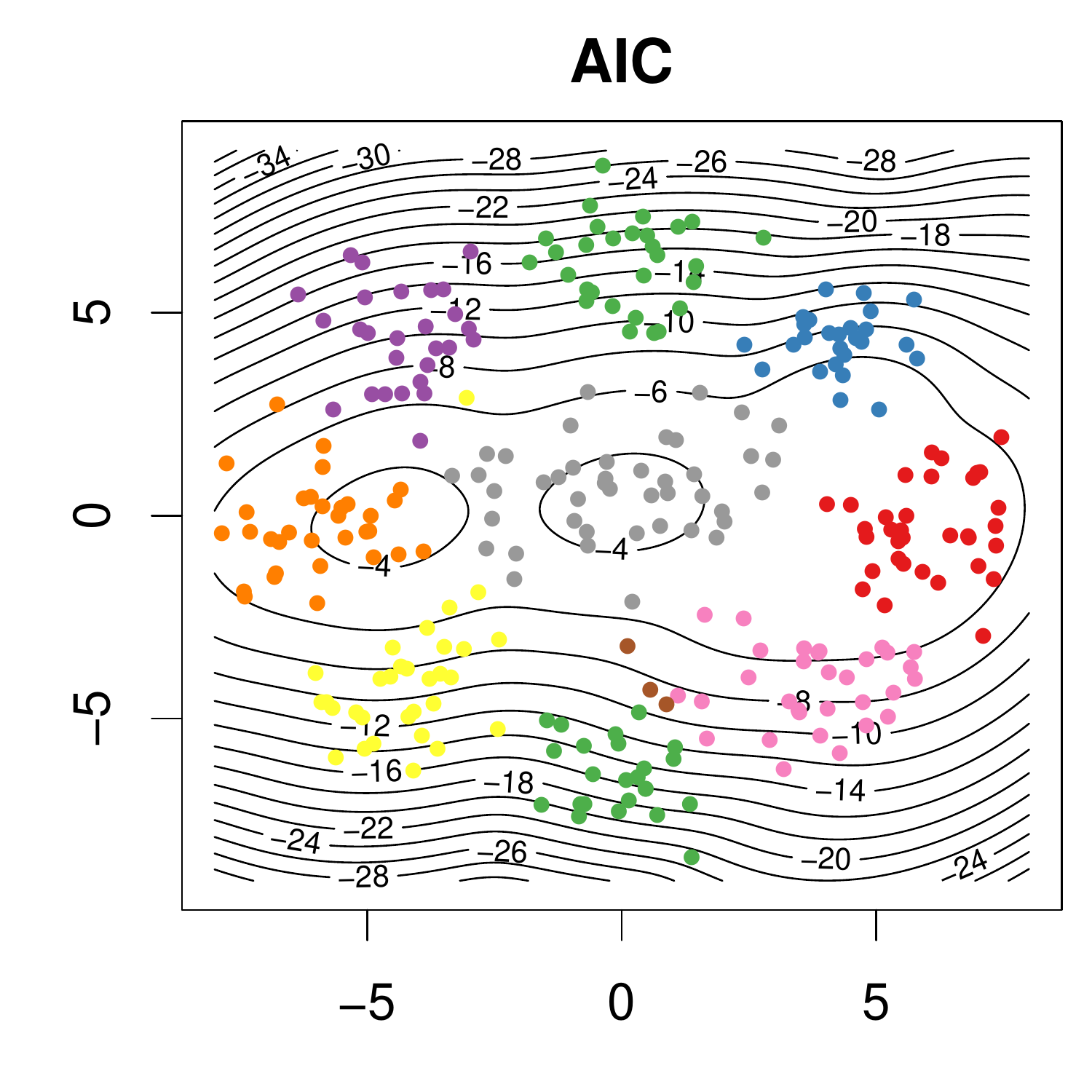}\\
\end{tabular}
\caption{Multivariate dataset 2 ($K=9$). Top: raw MCMC output of $(\mu_1,\mu_2)$ and reordered values according to \code{ecr} and \code{aic} algorithms for a randomly sampled subset of $500$ MCMC iterations. Bottom: True density and clusters of data and corresponding estimates to \code{ecr} and \code{aic} algorithms.}\label{fig:mvn2}
\end{figure}

The raw and reordered MCMC output for the means is displayed in Figures \ref{fig:mvn} and \ref{fig:mvn2} (top). Since most methods produced almost identical results, only \code{stephens} (for the first dataset) and \code{ecr} (for the second) are shown, along with \code{aic} which produced different results. The estimated density and single best clustering are displayed in Figures \ref{fig:mvn} and \ref{fig:mvn2} (bottom). The resulting estimates and single best clusterings reported by \code{aic} are in stark contrast with the rest of the methods due to the poor performance of the ordering constraint $\mu_{11}<\ldots<\mu_{1K}$. However, a reasonable performance is obtained for the user-defined permutations based on the ordering according to $\mu_{1k} - 2\mu_{2k}$, $k=1,\ldots,K$, as shown in Table \ref{tab:mvn}.

\section{Concluding remarks}\label{sec:discussion}

The \pkg{label.switching} package contains eight relabelling algorithms in order to deal with the problem of non-identifiability in MCMC outputs of mixtures of distributions or hidden Markov models. The input depends on each method, while most of them require information that usually is directly or easily available from the MCMC output. In case that the number of components is small then all algorithms can be applied. When $K$ is large, it is suggested to consider only methods that are optimized using the \code{lpSolve} routine for the solution of the assignment problem (\code{dataBased}, \code{ecr}, \code{ecr.iterative.1}, \code{ecr.iterative.2} and \code{stephens}). Furthermore, all possible simple ordering constraints can be directly applied using the \code{constraint = "ALL"} argument. In addition, the \code{"USER-PERM"} option allows the researcher to add new output and make direct comparisons with the available methods.

In practice, the number of components is rarely known. Within a Bayesian framework, $K$ can be estimated using either Bayes factor approaches \citep{chib,carlin} or trans-dimensional MCMC samplers such as the Reversible Jump MCMC algorithm of \cite{Green:95} \citep{Richardson:97,dellaportas2006multivariate,papastamoulis2009reversible} and the Birth-Death MCMC sampler of \citet{stephens2000}. In the first case, a separate MCMC sample for each possible value of $K$ is available, hence each one of them can be directly used as input to the \pkg{label.switching} package. In the latter case, the trans-dimensional MCMC sample should be partitioned to subsets for each distinct sampled value of $K$. Then, in order to make inference conditionally on a given $K$ using the \pkg{label.switching} package, the input should correspond to the relevant MCMC draws.

As far as we are concerned, there are no previous efforts for an integrated software of methods dealing with the label switching problem. Given the substantial field of applications of mixture and hidden Markov models as well as the need for making straightforward MCMC inference on complex posterior distributions, the \pkg{label.switching} package offers a handy post-processing supplementary tool towards this direction.
  
\section*{Acknowledgements}

The author wishes to thank an Associate Editor and two anonymous reviewers for their valuable recommendations regarding certain parts of the software and for their helpful comments that considerably improved the manuscript.


\bibliography{label-switching}

\end{document}